\documentclass[11pt, a4paper]{scrartcl} 

\usepackage[utf8]{inputenc}    
\usepackage[T1]{fontenc}       
\usepackage[english]{babel}    
\usepackage{gensymb}
\usepackage{lmodern}    
\usepackage[font=small,labelfont=bf]{caption}
\usepackage{subcaption}      
\usepackage{amsmath}                      
\usepackage{graphicx}         
\usepackage[left=2.5cm, right=2.5cm, top=3cm, bottom=3cm]{geometry} 
\usepackage{hyperref}
\hypersetup{
    colorlinks=true,
    linkcolor=black,
    urlcolor=black,
    citecolor=black,
}
\usepackage[table]{xcolor}
\definecolor{elemgray}{HTML}{EBEBEB}
\definecolor{connblue}{HTML}{0D79F2}
\definecolor{interyellow}{HTML}{E5F20D}
\colorlet{elemrow}{elemgray!60}
\colorlet{connrow}{connblue!15}
\colorlet{interrow}{interyellow!25}
\newcommand{\er}{\rowcolor{elemrow}\cellcolor{white}}
\newcommand{\cnr}{\rowcolor{connrow}\cellcolor{white}}
\newcommand{\ir}{\rowcolor{interrow}\cellcolor{white}}
\newcommand{\critfig}[1]{%
  \IfFileExists{Figures/#1.pdf}%
    {\includegraphics[height=14mm,width=\linewidth,keepaspectratio]{Figures/#1}}%
    {\fbox{\parbox[c][2mm][c]{0.85\linewidth}{\centering\tiny #1}}}}
       
\newcounter{crit}
\renewcommand{\thecrit}{(\alph{crit})} 
\newcommand{\crit}[1]{\refstepcounter{crit}\label{#1}\thecrit}

\usepackage{graphicx} 
\usepackage{tikz}
\usepackage{multirow}
\usepackage[
    backend=biber,      
    style=ieee,        
    sorting=none,       
    dashed=false,       
    isbn=false,         
    url=false,          
    doi=true            
]{biblatex}
\usepackage{makecell}

\usepackage{longtable} 
\newcolumntype{L}[1]{>{\raggedright\arraybackslash}m{#1}}

\usepackage{pdflscape}

\usepackage{tabularx}
\usepackage{array}
\usepackage{makecell}

\newcolumntype{C}[1]{>{\centering\arraybackslash}m{#1}}

\newcommand{\bbvariant}[3]{%
    \begin{minipage}[t]{0.18\textwidth}
        \centering
        \includegraphics[width=\textwidth]{Figures/#2.pdf}\\[-5pt]
        { #1}\\[0pt]
        {\footnotesize #3}
    \end{minipage}%
}
\newcommand{\bcvariant}[3]{%
    \begin{minipage}[t]{0.20\textwidth}
        \centering
        \includegraphics[width=\textwidth]{Figures/#2.pdf}\\[-2pt]
        { #1}\\[0pt]
        {\scriptsize #3}
    \end{minipage}%
}
\newcommand{\bnvariant}[3]{%
    \begin{minipage}[t]{0.28\textwidth}
        \centering
        \includegraphics[width=1\textwidth]{Figures/#2.pdf}\\[-2pt]
        { #1}\\[0pt]
        {\scriptsize #3}
    \end{minipage}%
}
\usepackage{threeparttable}
\usepackage{booktabs}
\usepackage[nointegrals]{wasysym}
\usepackage{multirow}
\usepackage{authblk}

\begin{document}
\title{JOINTS: Classification and analytical evaluation of detachable dry connections for reusable concrete elements}

\author[1]{Julius Fründt}\author[1]{Vladislav Gudžulić}\author[1]{Günther Meschke}\author[1,2,3]{\\Roger A. Sauer}\affil[1]{Institute for Structural Mechanics, Ruhr University Bochum, 44801 Bochum, Germany}\affil[2]{Department of Structural Mechanics, Gdansk University of Technology, 80-233 Gdansk, Poland}\affil[3]{Mechanical Engineering, Indian Institute of Technology Guwahati, Assam 781039, India}
\date{
}

\maketitle

\begin{abstract}
\noindent
This paper presents a systematic classification for detachable dry joints
suitable for the structural reuse of reinforced concrete elements,
together with an analytical evaluation of representative configurations.
The classification, named JOINTS and applicable to concrete connections beyond the reuse context, organizes the design space along five categories whose initials spell JOINT: the Juxtaposition (spatial arrangement of the paired elements), the Origin (their prior structural function), the Interface geometry, the Newly introduced materials at the joint, and the force-Transfer mechanism. A sixth category, the Score, completes the acronym and evaluates each configuration for sustainability, cost-efficiency, and mechanical performance.

From this design space, three representative configurations are evaluated
analytically under active post-tensioning, differing in both their joint
geometry (planar, stepped, and angled) and their tendon
inclination. The evaluation covers failure of the anchorage of the post-tensioning member, the contact limit of the interface, and the material limit of the concrete adjacent to the interface. Although complex joint geometries theoretically enhance interlocking,
their benefit is limited by the compressive-shear strength of the
concrete. The reinforcement of the reused elements may be damaged near the cut, and the splitting reinforcement that a conventional design would require is not guaranteed to exist. The present analysis therefore conservatively assumes that the concrete at the joint face alone resists the combined compression and shear, neglecting any residual contribution from the existing reinforcement. This analytical evaluation
represents a first step toward quantifying the score over a parametric
range of connections, here focusing on the mechanical criterion.

\vspace{6pt}
\end{abstract}

\vspace{12pt}


\section{Introduction}
With global usage exceeding 4 billion tonnes annually, cement accounts for 36\% of all construction-related $\text{CO}_2$ emissions \cite{Habert2020}. To achieve significant emissions reduction, a promising approach is to upcycle existing concrete structures by separating and refurbishing reinforced concrete elements for reuse in new structures. In \cite{Kuepfer2025} an Atlas of Reused Concrete (ARC) is proposed, giving an overview of the very few examples of global projects using reuse of concrete elements, whereby the potential for $\text{CO}_2$ and cost reduction is considerable. One reason for the low rate of construction with reused elements lies in the refurbishment and design of connections. The connection of reused concrete elements poses particular challenges compared to newly manufactured precast concrete connections. Another reason is that the aged reinforced concrete comes with unknown stiffness, reinforcement and degradation. Furthermore, in the context of circular reuse, easy dismantling, the reduction of newly applied materials and the limitations in the fine processing also pose challenges in the design of connections.
In segmental tunnel construction, lining is made possible by the use of precast concrete segment rings. The joint between the segments must be able to transmit high compressive forces and ensure the water tightness of the structure \cite{Putke2015}. In segmental bridge construction, on the other hand, prestressed dry and wet joints are used primarily to transfer shear forces \cite{Podolny1982}. In precast building construction, a wide variety of design approaches are available. The main property is a quick and easy assembly of dry, wet or glued joints \cite{Chang2023}. Furthermore, the use of ultra high performance concrete (UHPC) \cite{Fehling2014}, ultra high performance fiber reinforced concrete (UHPFRC) \cite{Oettler2023} or carbon-fiber reinforced polymer (CFRP) concrete \cite{Rettinger2023} allows for complex connection designs, like fine profiling of the contact surface, thereby increasing the structural performance of the connection.
Such dry joints have been classified and evaluated by Lanwer et al. \cite{Lanwer2022}, but for additively manufactured UHPC segments, with
criteria centered on manufacturability and assembly geometry and with
profiles shaped by CNC milling. New connection designs for precast structures have also been proposed, linking joint geometry to robotic manufacturing and re-utilization of the concrete elements \cite{baghdadi2024}.
With regard to the reuse of existing concrete elements, \textcite{Kuepfer2022} conducted a systematic review of already realized reuse projects, from which they derived and categorized the connection types documented therein, differentiating between those involving additional reinforcement and those involving additional concrete.
However, the review does not discuss how these connection types behave structurally, and what is their individual cost and environmental assessment. 

This work proposes a classification that does. It spans the connection design space rather than document case-studies, and it assesses cost and environmental performance qualitatively at the level of the individual connection rather than the whole project. In addition, the structural behavior of selected connection variants is treated analytically. The analysis identifies the governing limit states, the interactions between failure modes, and the points where deeper investigation is needed. Although focused on concrete reuse, the classification also applies to prefabricated concrete connections and, in principle, to connections in construction at large. The variants in each category remain open to extension.

The remainder of this article is structured as follows. Section \ref{sec:class} develops the classification framework, named JOINTS after the six categories that compose it. Section \ref{sec:structural_evaluation} then evaluates three representative post-tensioned configurations analytically -- a planar, a stepped, and an angled joint -- that differ in both interface geometry and tendon inclination. Conclusions are drawn in Section \ref{sec:discussion}.

\section{Classification of detachable connections}
\label{sec:class}

In circular construction, the structural system is synthesized from the specific geometries and properties of reused elements drawn from existing source structures. This paradigm, termed \emph{form follows availability} \cite{Gorgolewski2019}, inverts the conventional engineering logic: rather than deriving component dimensions from idealised structural demand, the existing element stock defines the feasible design space for the new load-bearing assembly \cite{Rose2025}.
\begin{figure}[h!]
    \centering
    \includegraphics[width=0.8\textwidth]{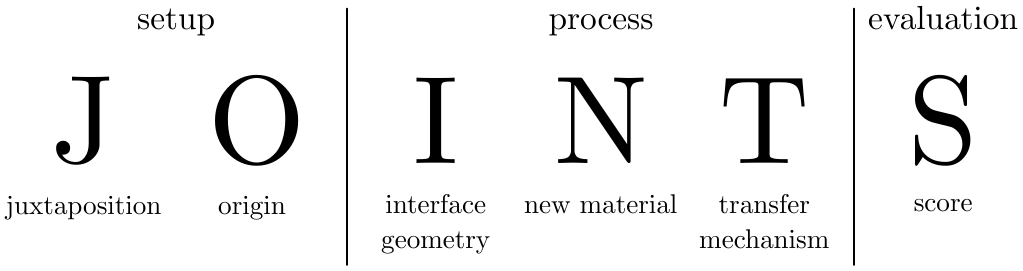}\\
    \caption{The \textsc{JOINTS} classification code. The six categories are
grouped into the setup of the connection (J, O), its process
(I, N, T), and its evaluation (S).}
    \label{fig:joints}
\end{figure}
The detachable dry connection variants identified in the succeeding sections are systematized under this paradigm along a six-letter classification code that spells \textsc{JOINTS}, see Figure \ref{fig:joints}. Five of the categories (\textbf{J}, \textbf{O}, \textbf{I}, \textbf{N}, \textbf{T}) are descriptive; the sixth, \textbf{S}, is an aggregated qualitative score. The six categories are grouped into three workflow phases that follow the design logic of reuse construction:

\smallskip
\noindent\textit{Setup} --- the boundary conditions defined by
structural intent and available inventory:
\begin{itemize}
    \item Section~\ref{sec:arrangement} introduces the \textbf{J}uxtaposition,
    that is, the spatial arrangement of the joined elements relative
    to each other.
    \item Section~\ref{sec:origin} specifies the \textbf{O}rigin of
    the pairing elements, classified by their prior structural
    function as beams, columns, plates, or wall slabs.
\end{itemize}

\smallskip
\noindent\textit{Process} --- the design decisions specifying how the
joint is constructed:
\begin{itemize}
    \item Section~\ref{sec:interface-geometry} addresses the
    \textbf{I}nterface geometry of the contact face.
    \item Section~\ref{sec:new-material} treats the \textbf{N}ewly
    introduced materials at the joint, covering all components added
    during reassembly.
    \item Section~\ref{sec:tensioning} describes the
    load \textbf{T}ransfer mechanism by which normal, shear, and
    moment actions are transferred across the dry interface.
\end{itemize}

\smallskip
\noindent\textit{Evaluation} --- the aggregated outcome:
\begin{itemize}
    \item Section~\ref{sec:score} synthesises the five preceding
    categories into a qualitative \textbf{S}core covering sustainability,
    cost-efficiency and mechanical performance.
\end{itemize}

\smallskip
Not all \textsc{JOINTS} combinations are mechanically consistent or feasible with aged reused elements with unknown reinforcement and tolerance. The score \textbf{S} reflects this restricted feasible design space.

A second restriction concerns detachability. The reused elements are themselves extracted from the source structure by sawing or cutting, and reproducing non-releasable joints in the new assembly would extend this destructive cycle into the next service life \cite{Mettke1995}. The classification therefore admits only connections with a releasable primary load path, one that can be deactivated by de-tensioning, unbolting, or unclamping so that the elements separate without damaging them. Wet joints and grouted connections, adhesive and chemical bonds, and welded steel inserts offer no such mechanism and are excluded, consistent with ISO~20887 \cite{ISO20887}, which identifies poured, welded, and chemically bonded connections as reducing the potential for disassembly. Within the admitted design space, detachability remains a matter of degree, ranging from fully reversible connections to configurations with sacrificial components in the element ends. This range is not an exclusion criterion but enters the score
\textbf{S} in Section~\ref{sec:score}. Although developed for reused elements, the proposed categorization also applies to connections of newly manufactured elements.
\subsection{Juxtaposition}
\label{sec:arrangement}
\begin{figure}[h!]
    \centering
    \bcvariant{EE}{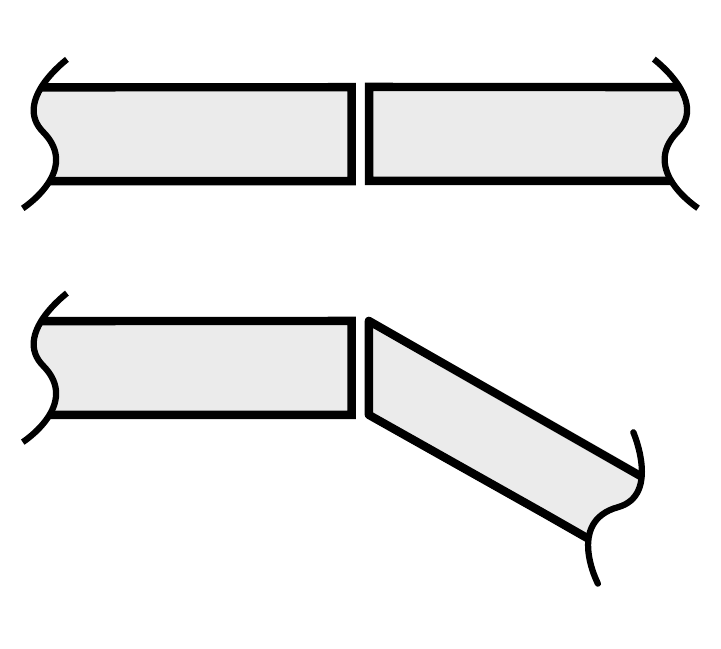}{End-to-End}%
    \hspace{5mm}
    \bcvariant{ES}{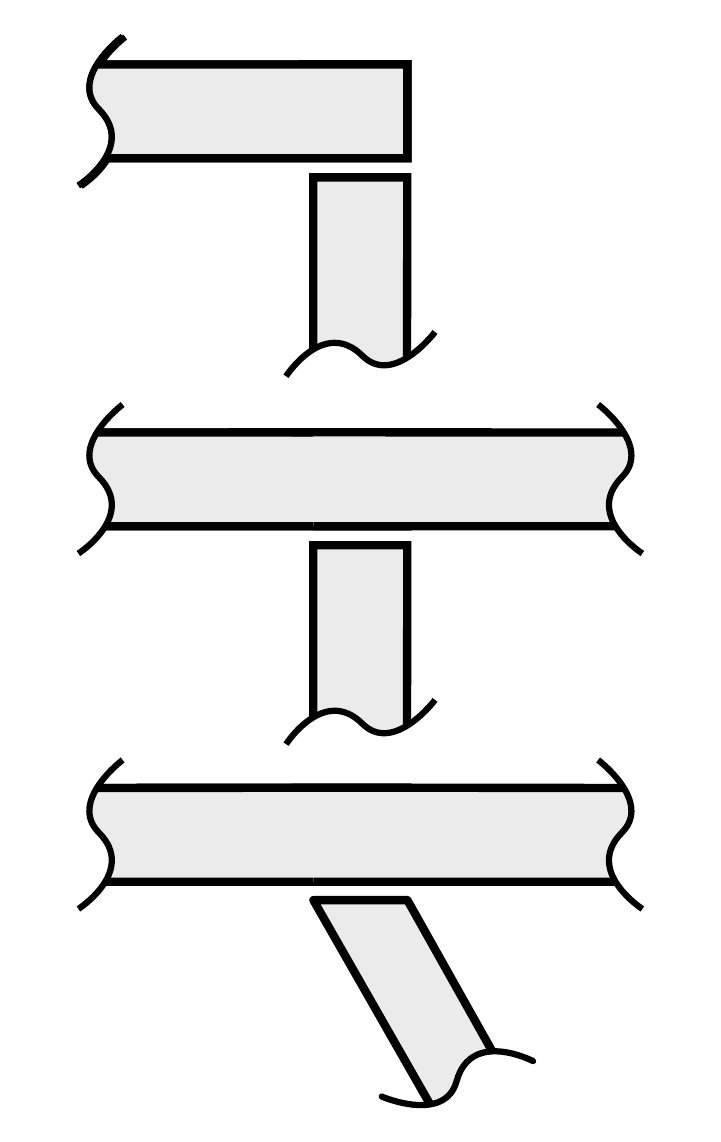}{End-to-Side}%
    \hspace{5mm}
    \bcvariant{SS}{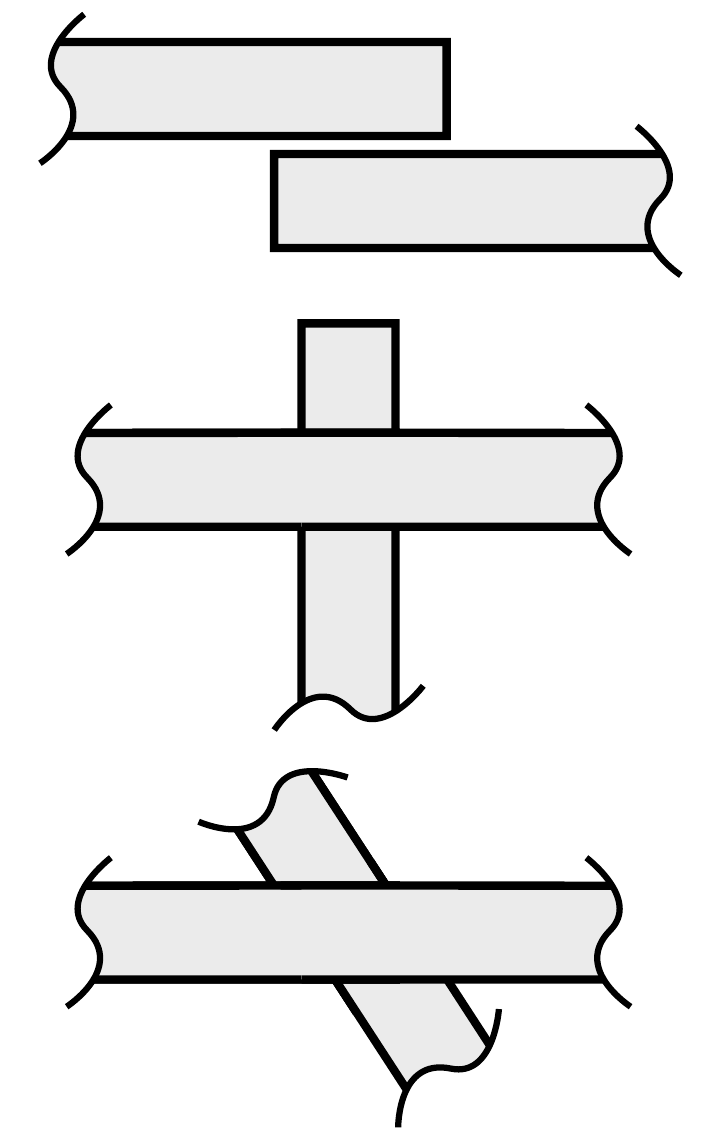}{Side-to-Side}%
    \caption{J: Element \textbf{J}uxtaposition.}
    \label{fig:top-catalogue}
\end{figure}

A first step in this design logic is to specify how reused elements are arranged relative to one another, independently of any specific connection detail, joint geometry, or auxiliary component. Every element is bounded by faces of two types: comparatively small end faces and larger side faces. For a beam or column these are the cross-sectional ends and the longitudinal sides. For a plate or wall slab, they are the edges and the two broad faces. Two elements brought into contact must meet through one of three possible pairings of these surface types, shown in Figure~\ref{fig:top-catalogue}. The End-to-End configuration places two end faces in contact. The End-to-Side configuration places an end face of one element against the side face of the other. The Side-to-Side configuration brings the side faces of both elements into contact. These three classes are exhaustive at the two-element level, since two surface types admit exactly three pairings. This reduction treats the edges of a planar element as equivalent. In the case of highly orthotropic elements, it may be necessary to distinguish further between the edge faces, but this will not be considered.
Multi-element junctions in a complete structural assembly arise as compositions of these primitives. A node at which three or four elements meet combines two or more of the three classes at a single point. The combinatorial space of such compositions is not enumerated in this work.
\subsection{Origin}
\label{sec:origin}
\begin{figure}[t!]
\centering
\begin{minipage}[c]{0.5\textwidth}
    \centering
    \includegraphics[width=\linewidth]{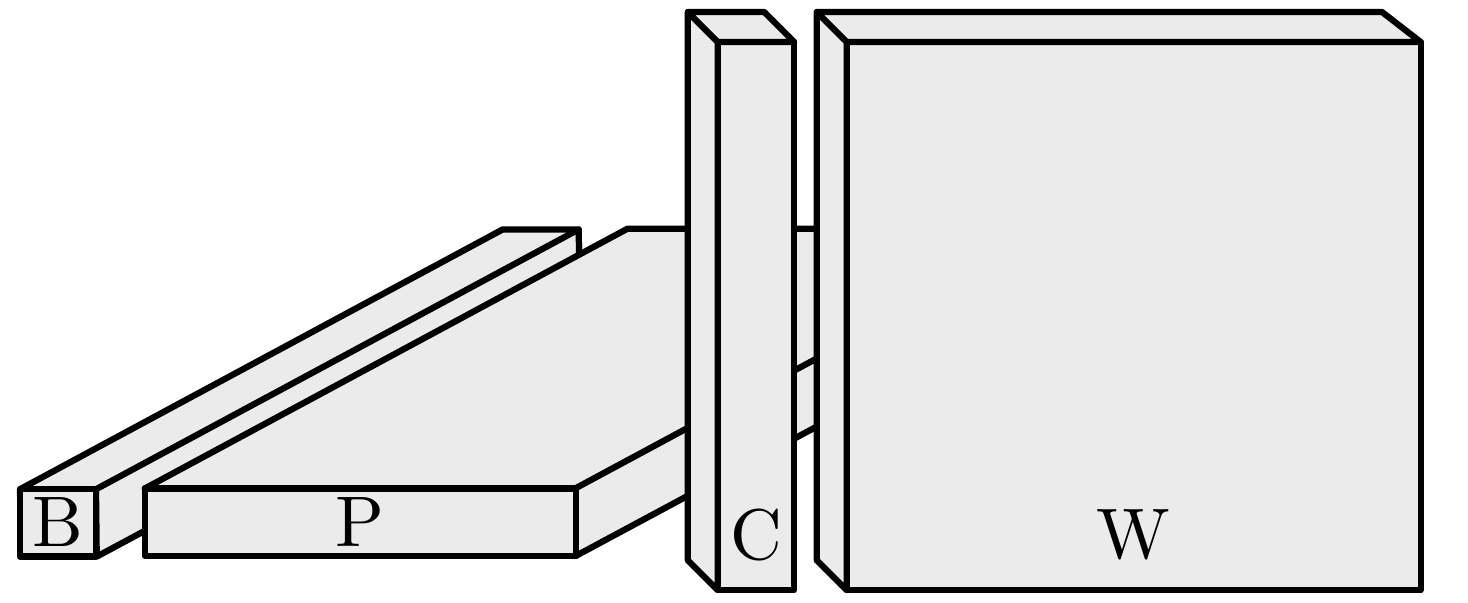}
\end{minipage}%
\hfill
\begin{minipage}[c]{0.45\textwidth}
    \centering
    \begin{tabular}{c@{\hspace{9mm}}l}
    \toprule
    \textbf{Code} & \textbf{Element pairs} \\
    \midrule
    BB & Beam -- Beam   \\[4pt]
    BC & Beam -- Column \\[4pt]
    BP & Beam -- Plate   \\[4pt]
    BW & Beam -- Wall slab   \\
    \vdots & \vdots    \\
    \bottomrule
    \end{tabular}
\end{minipage}
\caption{O: Element \textbf{O}rigin and corresponding code for pairing two elements.}
\label{fig:junction-partners}
\end{figure}
The \textit{origin} category determines the prior structural function of the two members entering the joint. An element's prior function as beam, column, plate, or wall slab corresponds to a characteristic reinforcement layout in the element. This constrains qualitatively which joining configurations are plausible for a given element. However it is not a basis for capacity verification. The cut severs the bars at the new joint face and removes their anchorage. The category value for a joint is the pair of the two members' prior functions. All ten pairs of the four element types are admissible, of which Figure~\ref{fig:junction-partners} lists representative beam-based examples.
The \textit{origin} category describes only the prior function of each member. It does not imply that the two members came from the same structure, nor does it describe how they are arranged in the new assembly. Two members that were beams in separate buildings, brought together for reuse, are classified as (BB) here regardless of their prior co-location or new use.
\subsection{Interface}
\label{sec:interface-geometry}
\begin{figure}[b!]
    \centering
    \bbvariant{P}{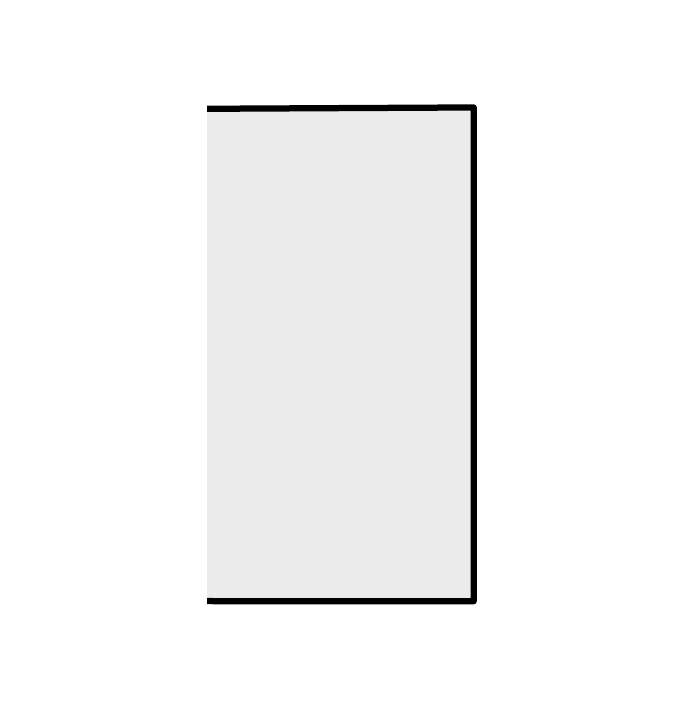}{Planar}%
    \hspace{2mm}
    \bbvariant{S}{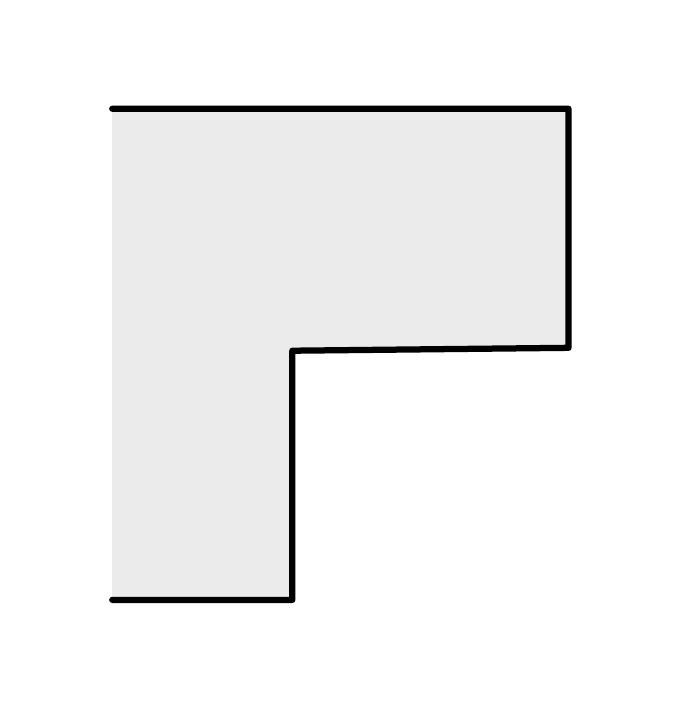}{Stepped}%
    \hspace{2mm}
    \bbvariant{A}{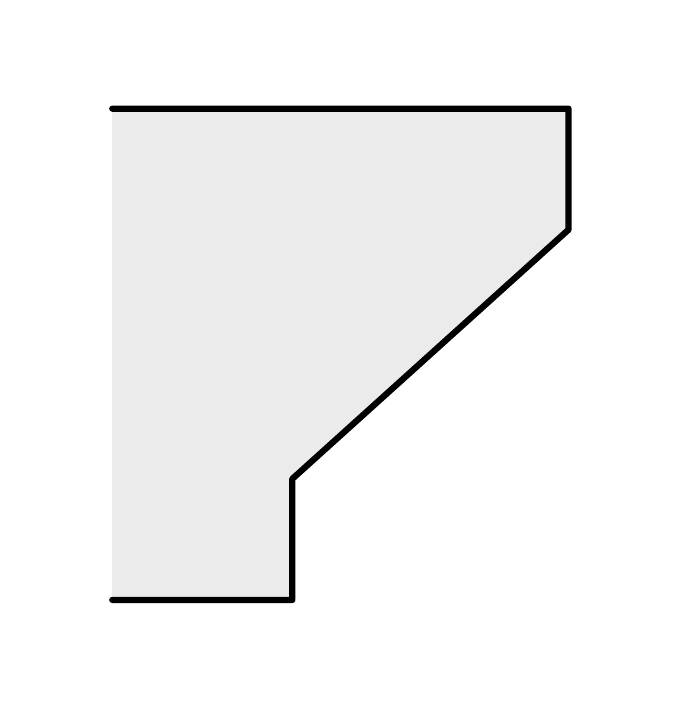}{Angled}%
    \hspace{2mm}
    \bbvariant{C}{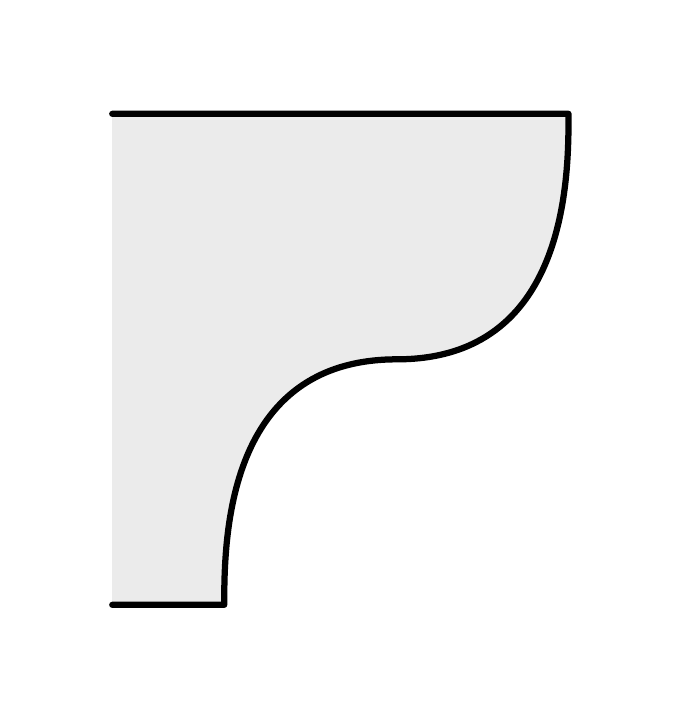}{Curved}%
    \\
    \bbvariant{K}{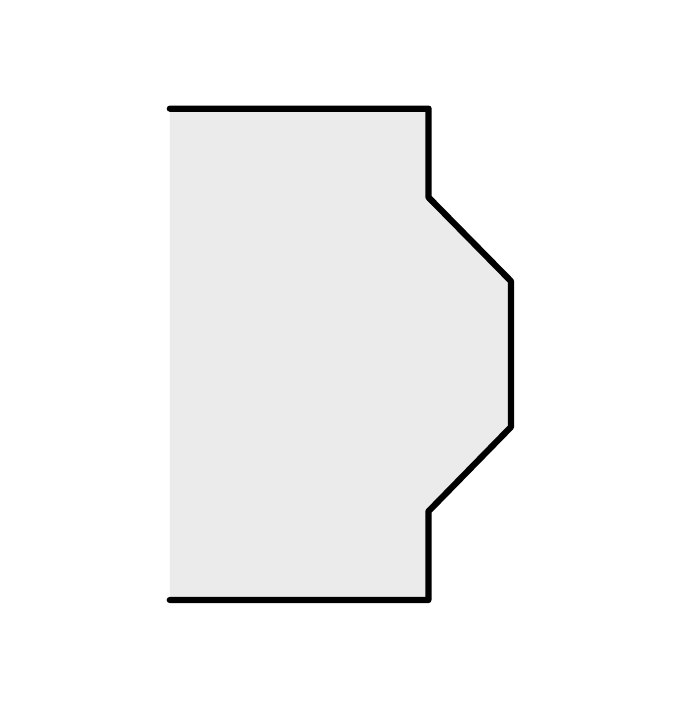}{Shear key}%
    \hspace{2mm}
    \bbvariant{T}{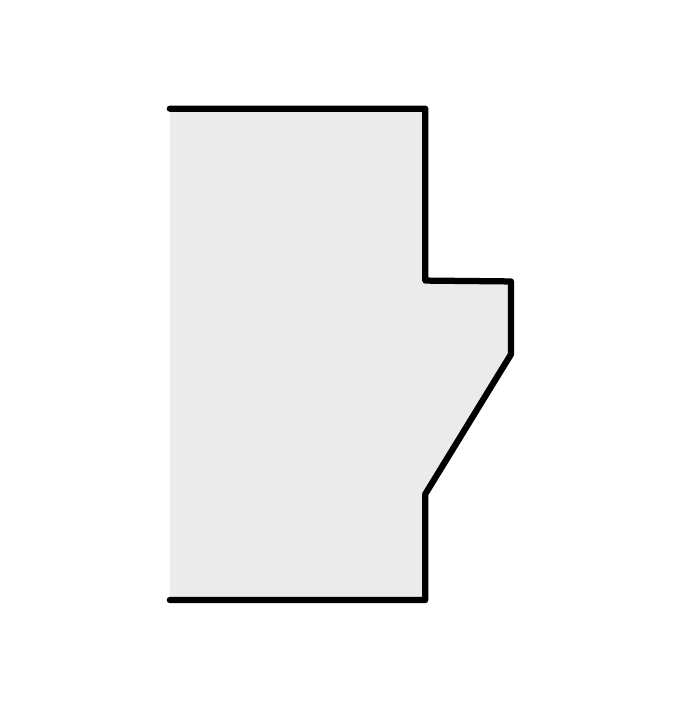}{Saw tooth}%
    \hspace{2mm}
    \bbvariant{W}{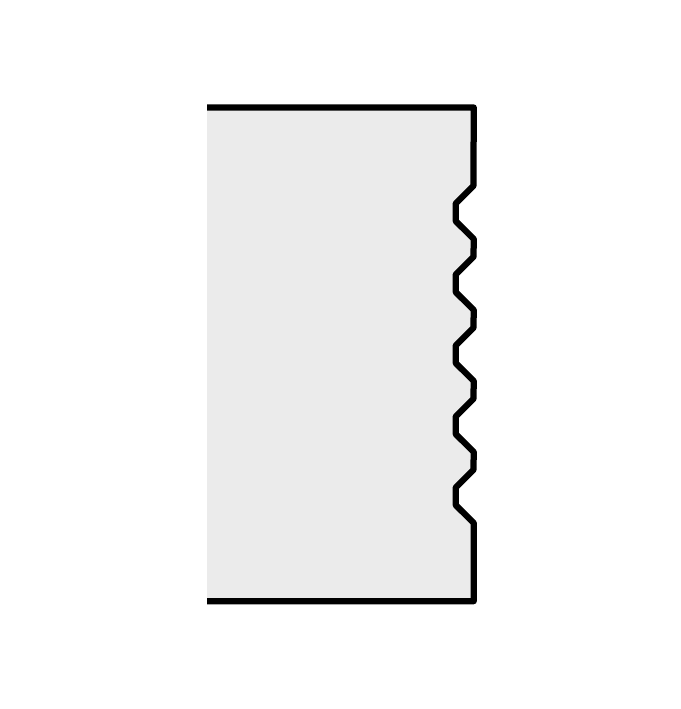}{Wavy}%
    \caption{I: Different \textbf{I}nterface geometries.}
    \label{fig:jg-catalogue}
\end{figure}
The \textit{interface geometry} of a joint strongly constrains the load-transfer mechanism, the manufacturing complexity, and the assembly kinematics of the joined elements. Two mechanisms act in parallel at the interface: geometric interlocking and friction. The first is caused by macroscale surface features, while the second is caused by microscale surface roughness \cite{Schaarschmidt2023}. Both increase the shear resistance of the joint. Interlocking has the additional advantage of self-alignment
during assembly.

This section classifies the macroscopic geometry through seven representative profiles, see Figure~\ref{fig:jg-catalogue}. The families form an open scheme to which further profiles can be assigned.

The planar interface (P) is characterized by a single planar segment with no transitions and is parametriz\-ed by the inclination of the interface. The stepped and angled interfaces (S and A) add one transition to an
otherwise planar profile, parametrized by the step size or angle, respectively. The stepped profile S is widely
established in precast construction for simply supported beam-beam or beam-column connections \cite{BetonVerlag1987}. The short vertical edges of the angled profile A at the top and bottom avoid acute concrete wedges at the element corners. The curved interface (C) comprises profiles with continuous curvature and no discrete transitions; their production requires curved cuts within the capabilities of water-jet or diamond-wire systems. The shear key interface (K) provides a bidirectional shear interlock, which is investigated in \cite{Stoehr2025}. The saw tooth interface (T), on the other hand, provides a unidirectional interlock. The wavy interface (W)
consists of a periodic sequence of transitions that distributes shear capacity across multiple interlock units, following principles
established in segmental bridge construction \cite{Guckenberger1980}. This interface geometry demands a dimensional accuracy beyond that of
the simpler geometric interfaces. In precast segmental construction, this accuracy is achieved by match-casting, in which the hardened
segment serves as formwork for the subsequent one \cite{Podolny1982}.
\subsection{New material}
\label{sec:new-material}
\begin{figure}[b!]
    \centering
\makebox[58mm][s]{Contact face 1\hfill Contact face 2}\\[2pt]
\includegraphics[width=50mm]{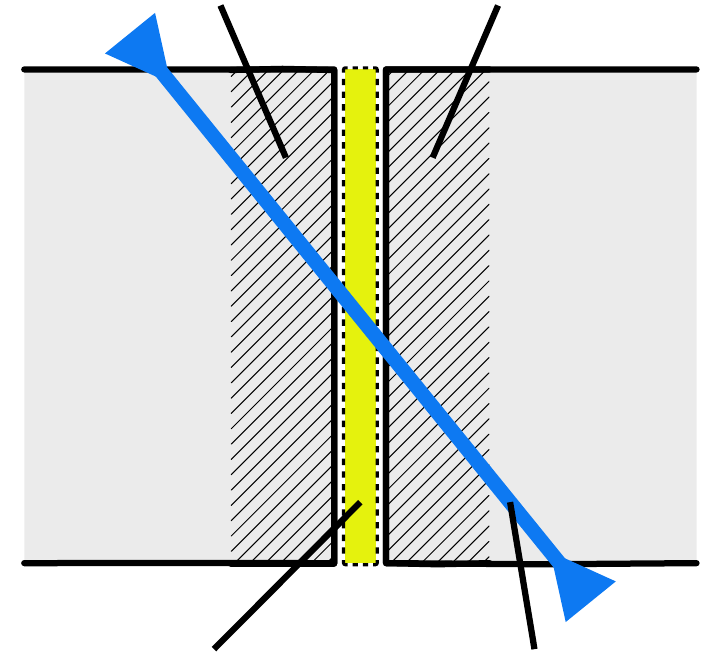}\\[-4pt]
\makebox[65mm][s]{\hspace{8mm} Interlayer\hfill Connecting material}
    \caption{N: \textbf{N}ew material added at the joint.}
    \label{fig:il-concept}
\end{figure}
Beyond the \textit{interface geometry}, the structural behavior of the contact zone depends on the \textit{new material} added at
the joint. As Figure~\ref{fig:il-concept} shows, this can be contact face material, interlayer material or connecting material. The added material can be classified as element-wise, specified for each pairing element
individually and possibly differing between the two, or joint-wise, specified once for the whole joint. Examples are given in Table~\ref{tab:new-material}.
\begin{table}[t!]
\centering
\setlength{\aboverulesep}{0pt}
\setlength{\belowrulesep}{0pt}
\renewcommand{\arraystretch}{1.15}
\begin{tabular}{c c l l}
\toprule
 & \textbf{Code} & \textbf{Description} & \textbf{Material property} \\
\midrule
\er & \multicolumn{3}{l}{\textit{Contact face material (bonded to each cut face)}} \\
\er & 0$-$0 & No added face material      & Original concrete face \\
\er & M$-$M & Mortar bed                  & High compressive strength \\
\er & F$-$F & Fiber reinforced mortar bed & High compressive and tensile strength \\
\er & S$-$S & Embedded steel plate        & Increased compressive strength \\
\cmidrule(l){2-4}
\er & \multicolumn{3}{l}{\textit{Contact face finish (Post-treatment on material face)}} \\
\er & $-_{\text{r}}$ & Roughened face     & Increased friction \\
\er\multirow{-8}{*}{\rotatebox{90}{\small element-wise}}
    & $-_{\text{g}}$ & Ground face        & Lowered friction \\
\cmidrule[\lightrulewidth]{1-4}
\cnr & \multicolumn{3}{l}{\textit{Connecting material}} \\
\cnr & $-$0$-$ & No connecting material & --- \\
\cnr & $-$S$-$ & Steel member           & Tensile and compressive strength, ductile \\
\cnr & $-$P$-$ & Fiber reinforced polymer strip & High tensile strength, brittle \\
\cnr & $-$K$-$ & Concrete shear key (loose) & High compressive strength, brittle \\
\cmidrule(l){2-4}
\ir & \multicolumn{3}{l}{\textit{Interlayer material}} \\
\ir & $-_{\text{0}}$ & No interlayer          & Direct dry contact \\
\ir & $-_{\text{F}}$ & Sliding foil           & Low friction coefficient \\
\ir & $-_{\text{L}}$ & Felt leveling pad      & Compressible, conforming \\
\ir & $-_{\text{E}}$ & Elastomeric bearing    & Elastic, viscoelastic damping \\
\ir\multirow{-11}{*}{\rotatebox{90}{\small joint-wise}}
    & $-_{\text{B}}$ & Steel-faced bearing    & High through-thickness stiffness \\
\bottomrule
\end{tabular}
\caption{N: \textbf{N}ew material added at the joint: Classification of added materials by primary material property. Face material is specified element-wise and may differ between the two pairing elements. Connecting and Interlayer materials are specified joint-wise.}
\label{tab:new-material}
\end{table}
\begin{figure}[b!]
    \centering
    \renewcommand{\arraystretch}{2.8}
   \begin{tabular}{@{}>{\centering\arraybackslash}m{0.25\textwidth}@{\hspace{1mm}}m{100mm}@{}}
        \begin{minipage}[c]{\linewidth}\centering
            \includegraphics[width=\linewidth]{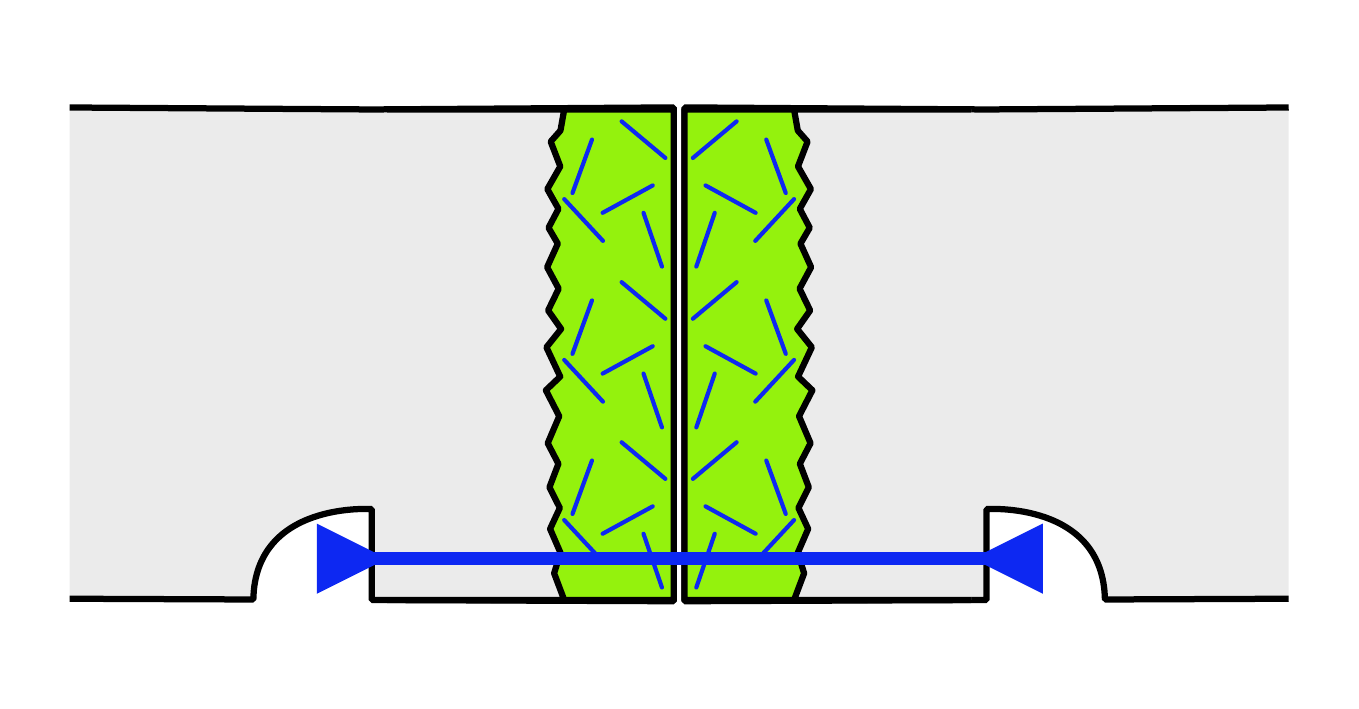}\\[-9pt]
            {$\text{F}_\text{g}\text{S}_\text{0}\text{F}_\text{g}$}
        \end{minipage}
        & Joint with a steel connecting member, with bonded fiber-reinforced mortar beds ground on both faces.
        \\[9mm]
        \hline
        \begin{minipage}[c]{\linewidth}\centering
            \includegraphics[width=\linewidth]{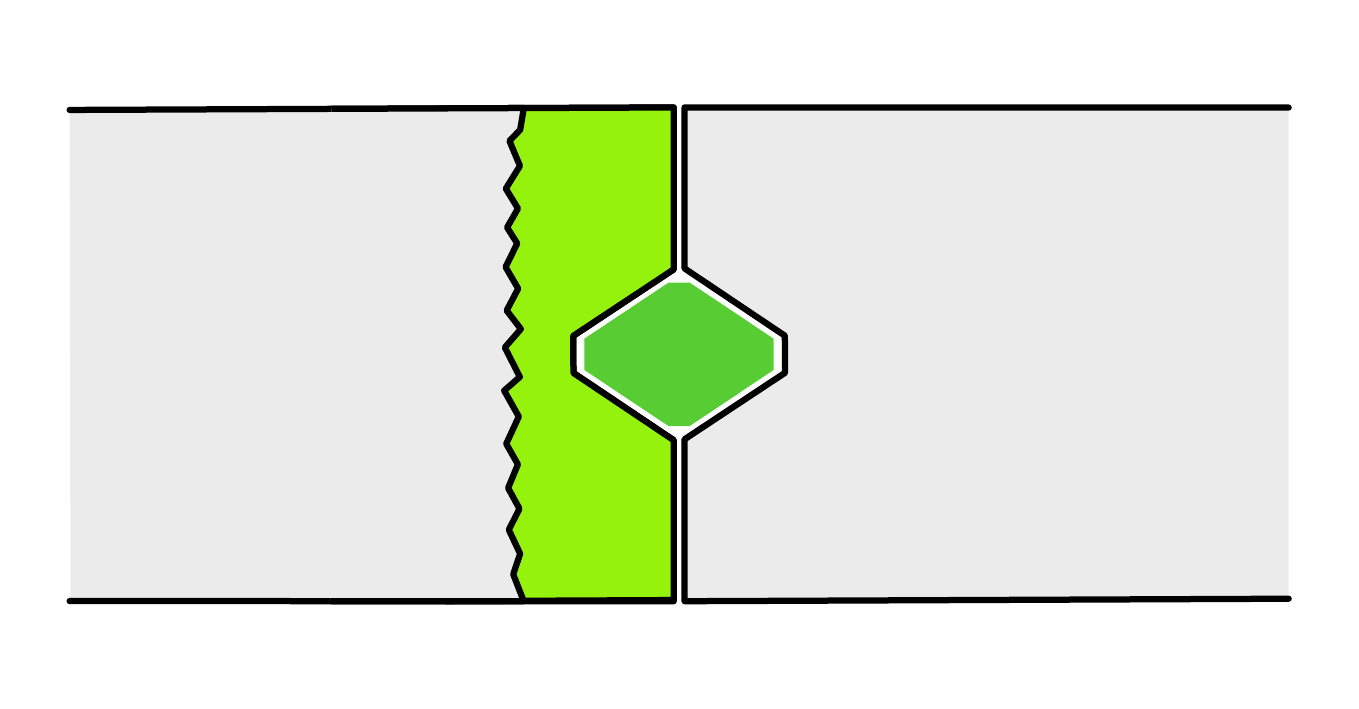}\\[-9pt]
            {$\text{M}_\text{g}\text{K}_\text{0}\text{0}_\text{g}$}
        \end{minipage}
        & Joint with a loose concrete shear key element bridging a bonded mortar bed and ground original concrete face.
        \\[9mm]
        \hline
        \begin{minipage}[c]{\linewidth}\centering
            \includegraphics[width=0.7\linewidth]{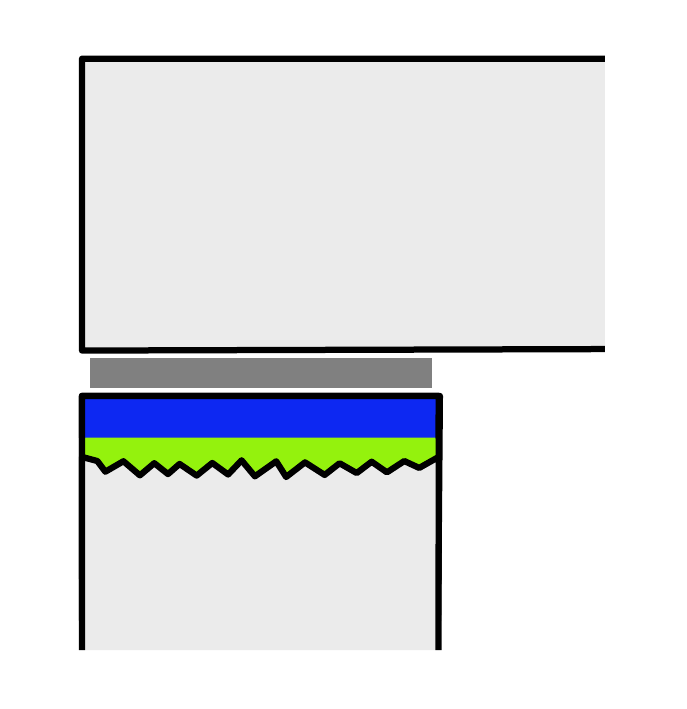}\\[-9pt]
            {$\text{0}_\text{r}\text{0}_\text{L}\text{S}_\text{g}$}
        \end{minipage}
        & Joint with felt levelling pad between a roughened original face and an embedded steel plate.
        \\
    \end{tabular}
\caption{N: \textbf{N}ew material added at the joint: Three example joint configurations with their corresponding \textbf{N}-category encoding.}
    \label{fig:new-material}
\end{figure}
Element-wise, each cut face carries a face material and a surface finish. The face material (0, M, F, S) specifies the composition of the contact face: the original concrete face after cutting with no added material (0), a bonded mortar bed (M) of high compressive strength, a fiber-reinforced mortar bed (F), which adds tensile capacity and
reduces local fracture sensitivity at corners with stress concentrations, or a bonded layer with an embedded steel plate (S),
which dominates the contact behavior as a stiff load-distribution element. In the bonded variants M, F, and S, the layer is applied to
the cut surface and becomes the new contact face after curing; it smooths the residual roughness of the original cut concrete and admits substantially higher contact pressures, the achievable level being set by the layer material rather than by the cut concrete. The finish is marked by a subscript, roughened (r) or ground (g): a roughened finish increases friction and follows from a
mechanical treatment such as sandblasting, whereas a ground finish lowers friction and arises from a slow sawing process \cite{GrosseKreul2025} or from a subsequent polishing operation.
Form deviations on the component scale are not classified here. They are, however, taken into account in mechanical evaluations described in Section \ref{sec:structural_evaluation}.
Joint-wise, the classification specifies a connecting material and an
optional interlayer. The connecting material (0, S, P, K) enhances force transfer across the interface: none (0), a steel member (S) of ductile tensile and compressive capacity, a fiber-reinforced polymer strip (P) of high but brittle tensile strength, or a loose concrete shear key (K) seated in matching recesses. The loose key (K) is distinct from a key cut into the \textit{interface geometry} (Section~\ref{sec:interface-geometry}): here it is an added component, not a feature of the face. The interlayer material, marked as a subscript on the connecting material encoding, is distinguished by mechanical function. Direct
dry contact carries none (0). A compliant felt leveling pad (L) or elastomeric bearing (E) accommodate residual rotation and distribute contact pressures without opening the joint,
whereas a steel-faced bearing (B) provides high stiffness. A thin sliding foil (F) serves the opposite purpose, deliberately decoupling the faces tangentially to
suppress friction transfer \cite{BetonVerlag1987}.

A complete \textbf{N}-code is a three symbol code with face material in the outer slots and connecting material in the
central slot. Each symbol carries a subscript to designate face finish and interlayer material. Figure~\ref{fig:new-material} encodes three exemplary configurations,
two beam--beam connections and one beam--column connection, which differ in the new material added joint- and element-wise.
\subsection{Transfer mechanism}
\label{sec:tensioning}
\begin{figure}[t!]
    \centering
    \bnvariant{0}{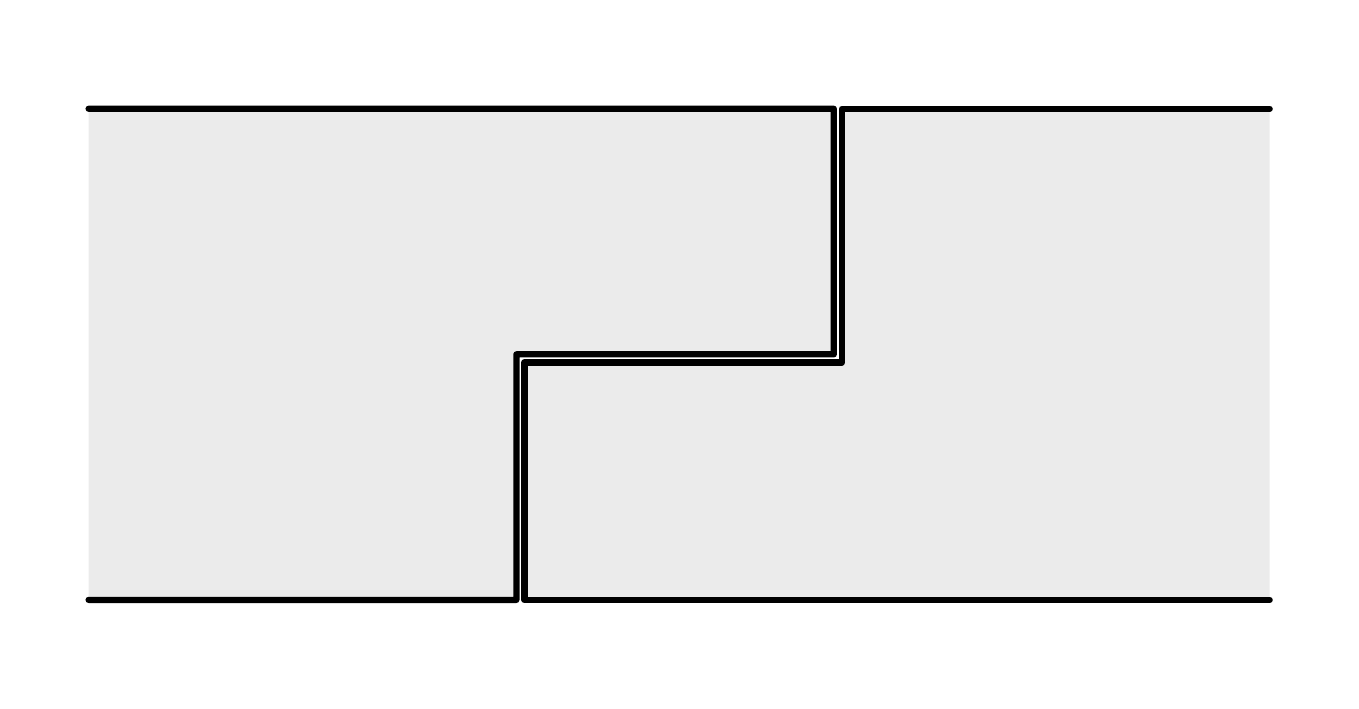}{Force transfer via interface friction and geometric interlocking}%
    \hspace{4mm}
    \bnvariant{BPB}{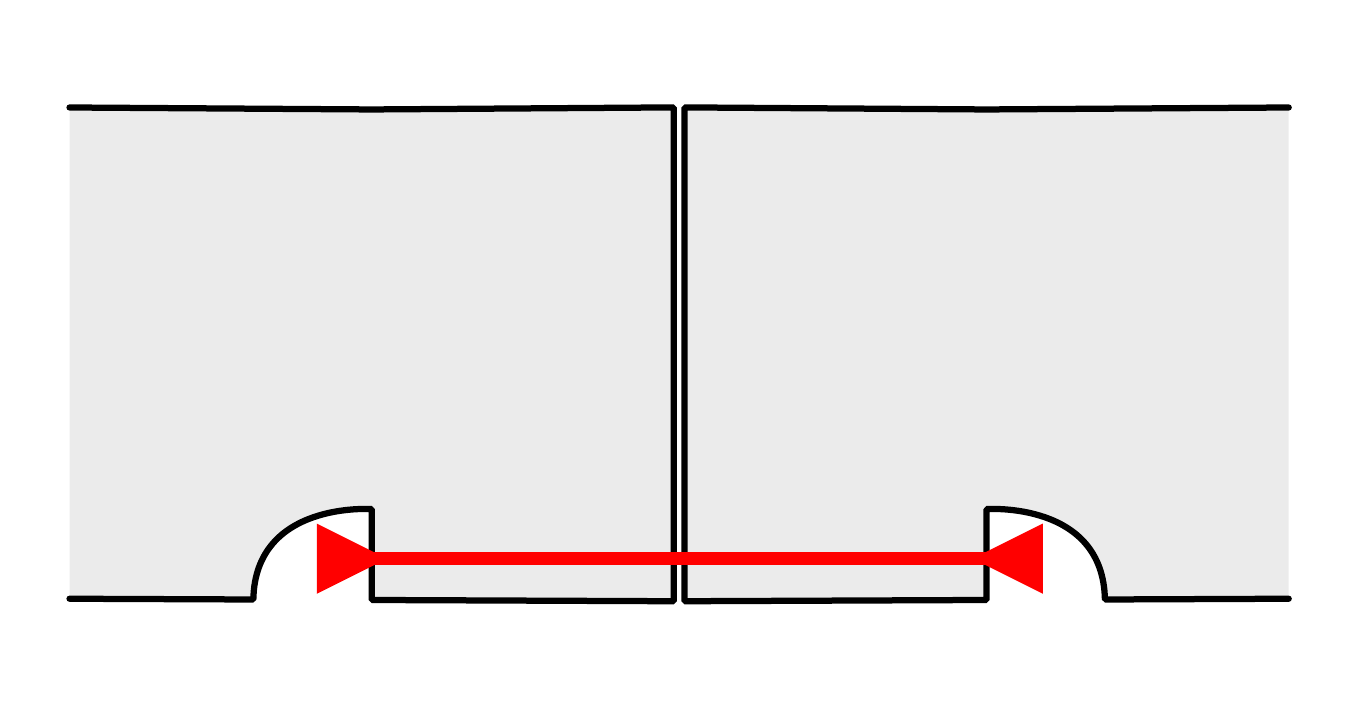}{Post-tensioned rod with bolted anchorage on both sides}%
    \hspace{4mm}
    \bnvariant{BTF}{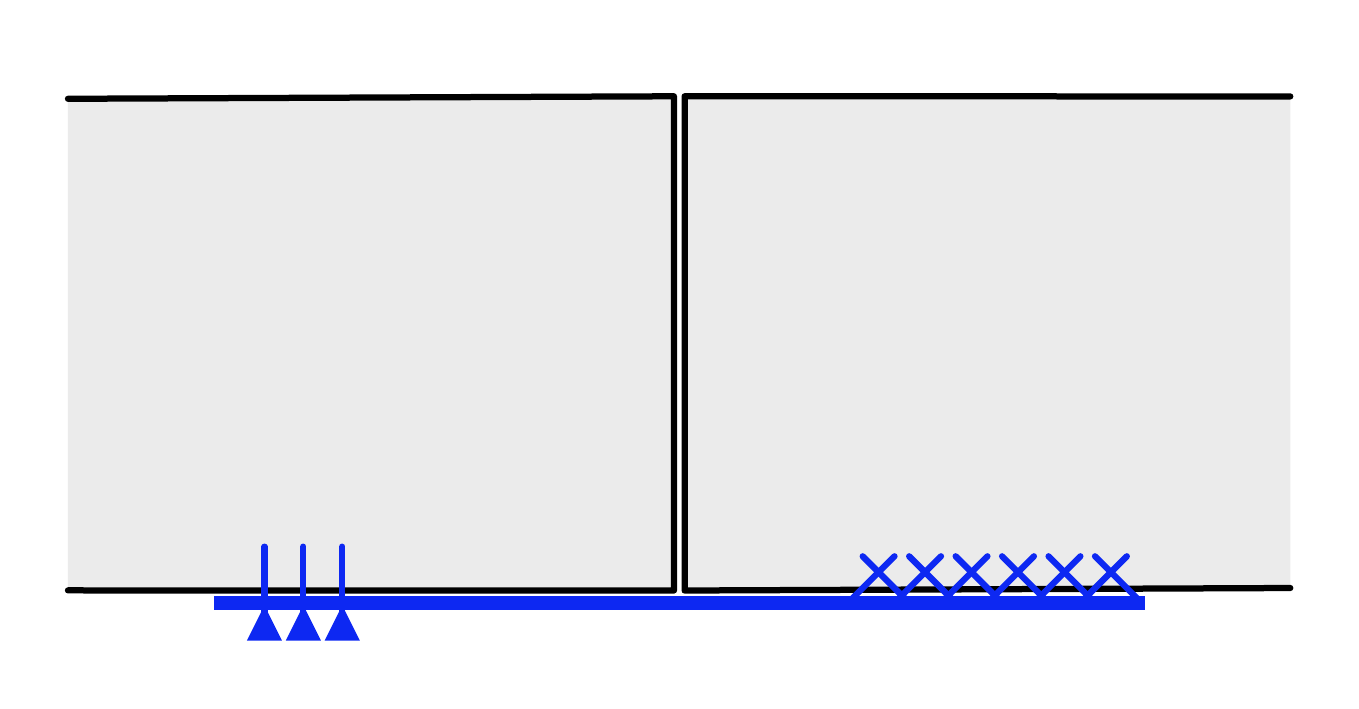}{Tensile steel tie, bolted at one end and grouted at the other}%
    \\[3mm]
    \bnvariant{CSC}{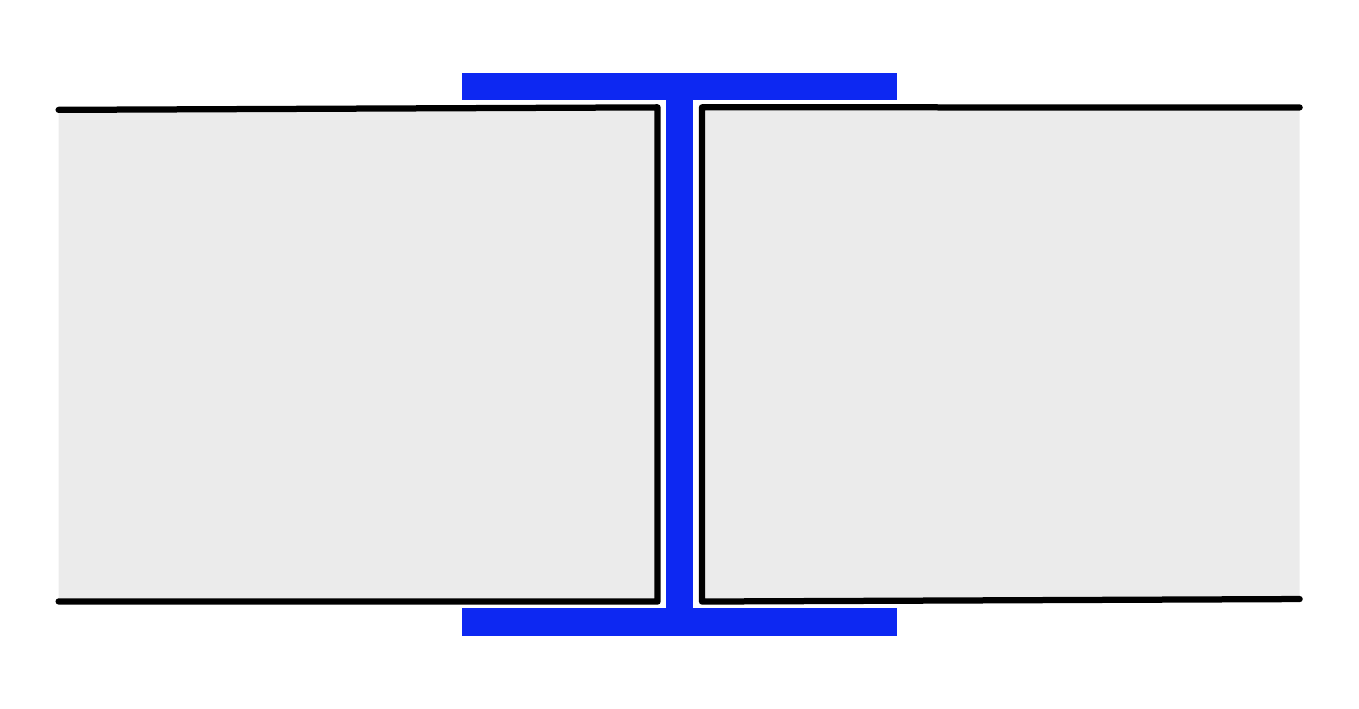}{Self-clamped steel profile transferring shear}%
    \hspace{4mm}
    \bnvariant{FMF}{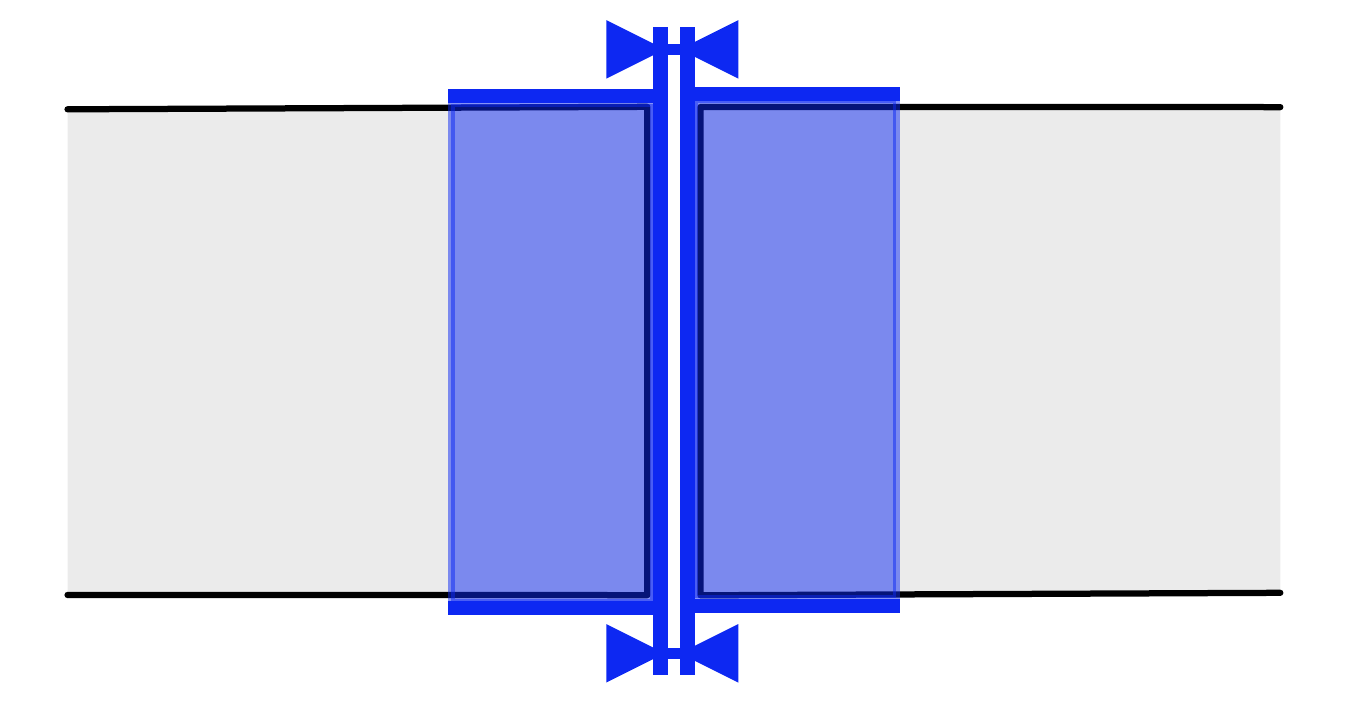}{Grouted steel housings linked by a releasable coupler}%
\caption{T: Force-\textbf{T}ransfer: Exemplary connections specified by force-transfer mode and attachment type.}
    \label{fig:transfer-mechanism}
\end{figure}

\begin{table}[b!]
\centering
\begin{threeparttable}
\begin{tabular}{c@{\hspace{4mm}} c l l}
\toprule
 & \textbf{Code} & \textbf{Description} & \textbf{Mechanical role} \\
\midrule
\multirow{6}{*}{\rotatebox{90}{\small joint-wise}}
  & \multicolumn{3}{@{}l}{\textit{Force transfer mode}} \\
  & $-$0$-$ & No connecting member  & Force transfer via interface only \\
  & $-$P$-$ & Post-tensioner        & Active axial pre-compression \\
  & $-$T$-$ & Tensile               & Passive tension transfer\\
  & $-$S$-$ & Shear                 & Passive shear transfer \\
  & $-$M$-$ & Multi-directional     & Combined axial, shear, and bending \\
\midrule
\multirow{4}{*}{\rotatebox{90}{\small element-wise}}
  & \multicolumn{3}{@{}l}{\textit{Attachment (per element side)}} \\
  & F$-$F & Permanently fixed\tnote{a} & Irreversible bond, e.g.\ grout or adhesive \\
  & B$-$B & Bolted          & Reversible mechanical fastening \\
  & C$-$C & Self-clamped    & Frictional clamping via contact pressure \\
\bottomrule
\end{tabular}
\begin{tablenotes}
\small
\item[a] Symmetric F$-$F configurations require a releasable coupler within the connecting member to preserve joint detachability, e.g. a turnbuckle.
\end{tablenotes}
\end{threeparttable}
\caption{T: Force-\textbf{T}ransfer classification: The force-transfer mode is specified joint-wise. The attachment is specified element-wise and may differ between the two sides of the connecting member.}
\label{tab:transfer-mechanism}
\end{table}
The final descriptive category addresses how the two joined elements are mechanically connected across the dry interface. Transfer may occur through the interface alone or through a connecting member spanning it. The classification presented in Table~\ref{tab:transfer-mechanism} is organized along two subcategories: the joint-wise force transfer mode and, where a connecting member is present, how it is attached to each element. Examples are shown in Figure~\ref{fig:transfer-mechanism}.
Without a connecting member (0), forces cross the joint through the interface alone, by compression, friction and geometric interlocking. The capacity is then set entirely by the \textit{interface geometry} of Section~\ref{sec:interface-geometry} and the contact materials of Section~\ref{sec:new-material}. The post-tensioner (P) is the only active mode: the member is pre-stressed during installation and applies a permanent axial compression to the joint, independent of subsequent loading. The remaining modes are passive and engage only as the joint deforms. A tensile member (T) resists axial separation through in-plane axial stiffness, idealised as
carrying axial tension only. Its bending stiffness is intentionally low, so that joint sliding induces member deflection rather than shear transfer. A shear member (S) resists in-plane sliding through transverse load transfer. A multi-directional member (M) carries several components simultaneously,
axial and shear through stiffer plate geometries, full encasing hardware, or volumetric reinforcement of the joint region.
The attachment category distinguishes three principles by which a connecting member is fixed to each element, ordered from least to most reusable and
thereby reflecting the central design criterion of this classification. A permanently fixed attachment (F) bonds the member to the element by grouting, embedding, or adhesion. Where both attachments are permanently fixed, detachability is preserved only if the member itself contains a releasable discontinuity within the joint region, typically a turnbuckle coupling
\cite{Santner2016}. A bolted attachment (B) connects the member through threaded fasteners through a bore hole. The member can be unscrewed and exchanged, but the drill holes and any inserts constrain the element's subsequent reuse
configurations. A self-clamped attachment (C) holds the member by
external contact pressure alone, without drilling, grouting, or bonding.
The element remains geometrically and materially unmodified, retaining
the broadest spectrum of subsequent reuse.

The force transfer code uses up to three letters to encode the force mode and the element-side attachments, which may differ between the two sides. The member-free baseline is encoded as 0 alone. For example, BPB denotes a post-tensioner bolted on both sides and BTF a tensile member bolted at one end and fixed at the other. The examples in Figure~\ref{fig:transfer-mechanism} span the five modes.
\subsection{Score}
\label{sec:score}
The preceding classification JOINT determines the connection and has evaluative implications. The \textit{score} \textbf{S} aggregates these into a composite evaluation for three criteria: sustainability, cost-efficiency and mechanical performance. A connection that prioritizes full detachability but uses a substantial amount of steel for its connecting members, or that excludes the elements from further reuse cycles, undermines the sustainability of reusing concrete elements. This section introduces the evaluation criteria in qualitative terms.

The evaluation of sustainability considers two aspects together. The first is the material, energy and carbon footprint of the connection. It comprises the steel and the mortar as well as the manufacturing and processing energy of the assembly required for the connection. In the reuse context, this footprint extends to the deconstruction of the elements from the source structure and their transport to the receiving structure. Relevant indicators include the quantity of steel in connecting members and embedded plates, the quantity of mortar or grout used in interface preparation, the energy required for cutting, drilling, and grinding of the element faces and the transport effort between source and receiving structure governed by element mass and distance. 
The second aspect of sustainability is the reusability of the elements after the connection has been assembled. A connection that prevents further reuse of the elements only postpones their demolition and
replacement. 
Reusability depends on the modifications the connection leaves in the elements and on attachments left by the connecting member. 
A non-planar interface requires a matching counterpart in the
next connection. Drilled ducts, holes, and recessed pockets are designed for one connection type, and may be unsuitable for other connection types. A self-clamping configuration leaves the elements unmodified. Whether the connecting member is self-clamped, bolted, or permanently fixed decides whether element and connecting member can be reused separately. High sustainability therefore
requires both a low immediate footprint and high reusability.

The evaluation of cost-efficiency captures the resources required to produce and assemble the connection. Relevant indicators are the
manufacturing time, the number of work hours, the energy consumption of cutting and processing, the cost of newly introduced components and the cost for transport of the elements.

The evaluation of mechanical performance considers the capacity of the connection to transfer section forces and its behavior at and beyond the design load. In the post-elastic range, the evaluation distinguishes between ductile failure with load redistribution and brittle failure with sudden loss of load bearing capacity. Further indicators are the stiffness of the connection, robustness against unforeseen loading, and durability under cyclic loading.

Section~\ref{sec:structural_evaluation} provides a first evaluation of selected connection variants for the three score criteria, with the main
focus on the mechanical criterion, which is evaluated quantitatively. A complete scoring across all three criteria and the full design space remains the subject of subsequent studies, for which the present classification provides the structure.

\section{Evaluation of connection variants}
\label{sec:structural_evaluation}
The evaluation proceeds in two steps. Section~\ref{sec:selection} evaluates six connection variants qualitatively with respect to sustainability and cost-efficiency and selects three of them. The mechanical performance of the three selected variants is then evaluated analytically using engineering criteria. 
Section~\ref{sec:procedure} sets out the procedure of the mechanical evaluation, 
Section~\ref{sec:anchorage_failure} treats failure at the anchorage of the post-tensioning member and Section~\ref{sec:interface_failure} covers the failure at the interface of the elements.
\subsection{Qualitative evaluation and variant selection}
\label{sec:selection}
To demonstrate how the JOINT classification affects the three criteria of the \textit{score} \textbf{S}, six connection variants are adapted from the literature and evaluated.
The variants differ only in the \textit{interface} (\textbf{I}) and the \textit{transfer mechanism} (\textbf{T}), while the \textit{juxtaposition}, \textit{origin}, and
\textit{new material} classifications (\textbf{J}, \textbf{O}, \textbf{N}) are held fixed at a baseline configuration. For the \textit{juxtaposition}, the End-to-End configuration (EE) is chosen because of its symmetry. Two cut ends are
joined, so both elements receive the same cut and the same anchorage detail. 
For the \textit{origin}, the element pair is set to Beam-Beam (BB). 
Since beams are designed to resist bending and shear, the connection between them must be capable of transferring both. This makes the pair suited to
demonstrate the evaluation of mechanical performance. In all variants, the only \textit{new material} is the steel of the connecting member (0S0). No new material is added to modify the properties of the element ends.

\begin{figure}[b!]
    \begin{tikzpicture}[
        x=2cm, y=2cm,
        marker/.style={draw, fill=black, inner sep=0pt, minimum size=7pt, line width=0.6pt},
        v label/.style={font=\small, inner sep=2pt},
        code/.style={font=\scriptsize, anchor=south west, inner sep=0pt}
    ]
    
    \draw[gray!25, thin] (1, 1.5) -- (4, 1.5);
    \draw[gray!25, thin] (1, 2.5) -- (4, 2.5);
    \draw[gray!25, thin] (1, 3.5) -- (4, 3.5);
    \draw[gray!25, thin] (1.5, 1) -- (1.5, 4);
    \draw[gray!25, thin] (2.5, 1) -- (2.5, 4);
    \draw[gray!25, thin] (3.5, 1) -- (3.5, 4);
    \draw[->, thick] (0.9, 1) -- (4, 1);
    \draw[->, thick] (1, 0.9) -- (1, 4);

    \node[font=\small, anchor=north] at (1.5, 0.95) {low};
    \node[font=\small, anchor=north] at (2.5, 0.95) {moderate};
    \node[font=\small, anchor=north] at (3.5, 0.95) {high};
    \node[font=\small, anchor=north] at (2.5, 0.65) {Sustainability};

    \node[font=\small, anchor=east] at (1, 1.5) {low};
    \node[font=\small, anchor=east] at (1, 2.5) {moderate};
    \node[font=\small, anchor=east] at (1, 3.5) {high};
    \node[font=\small, rotate=90, anchor=south] at (0.28, 2.5) {Cost-efficiency};

    \node[marker, circle] (v1) at (3.4, 3.0) {};
    \node[v label, anchor=west] at (v1.east) {1};

    \node[marker, circle] (v2) at (2.8, 2.5) {};
    \node[v label, anchor=west] at (v2.east) {2};

    \node[marker, circle] (v3) at (2.5, 2.1) {};
    \node[v label, anchor=west] at (v3.east) {3}; 

    \node[marker, circle] (v4) at (2, 2.5) {};
    \node[v label, anchor=west] at (v4.east) {4}; 

    \node[marker, circle] (v5) at (3.1, 3.4) {};
    \node[v label, anchor=west] at (v5.east) {5};

    \node[marker, circle] (v6) at (1.5, 1.5) {};
    \node[v label, anchor=west] at (v6.east) {6};

\begin{scope}[shift={(4, 2)}]
    \node[anchor=north west, inner sep=0pt] (v1img) at (0.3, 1.8)
        {\includegraphics[width=0.2\linewidth]{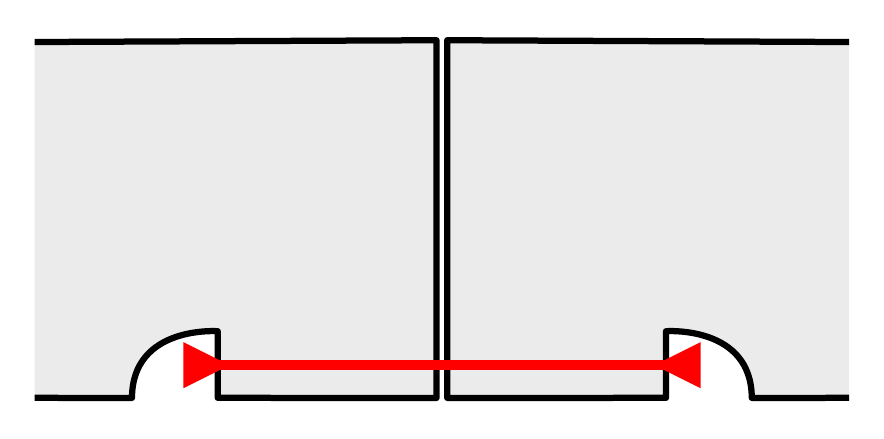}};
    \node[code] at ([xshift=0pt, yshift=-3pt]v1img.north west)
        {\begin{tabular}{@{}r@{}c*{5}{@{}c@{}c}@{}}
     & \textbf{J} & . & \textbf{O} & . & \textbf{I} & . & \textbf{N} & . & \textbf{T} & . & \textbf{S} \\
  1. & EE         & . & BB         & . & P          & . & 0S0        & . & BPB        & . & -
\end{tabular}};

    \node[anchor=north west, inner sep=0pt] (v2img)
        at ([yshift=-30pt]v1img.south west)
        {\includegraphics[width=0.2\linewidth]{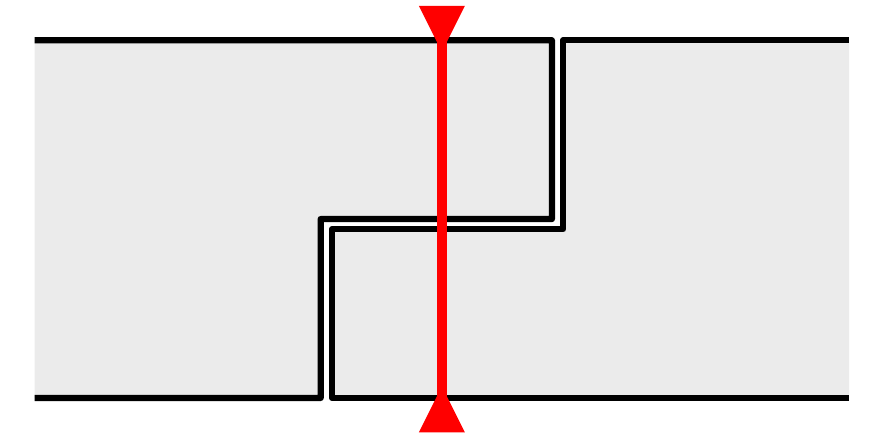}};
    \node[code] at ([xshift=0pt, yshift=-3pt]v2img.north west)
        {\begin{tabular}{@{}r@{}c*{5}{@{}c@{}c}@{}}
     & \textbf{J} & . & \textbf{O} & . & \textbf{I} & . & \textbf{N} & . & \textbf{T} & . & \textbf{S} \\
  2. & EE         & . & BB         & . & S          & . & 0S0        & . & BPB        & . & -
\end{tabular}};

    \node[anchor=north west, inner sep=0pt] (v3img)
        at ([yshift=-30pt]v2img.south west)
        {\includegraphics[width=0.2\linewidth]{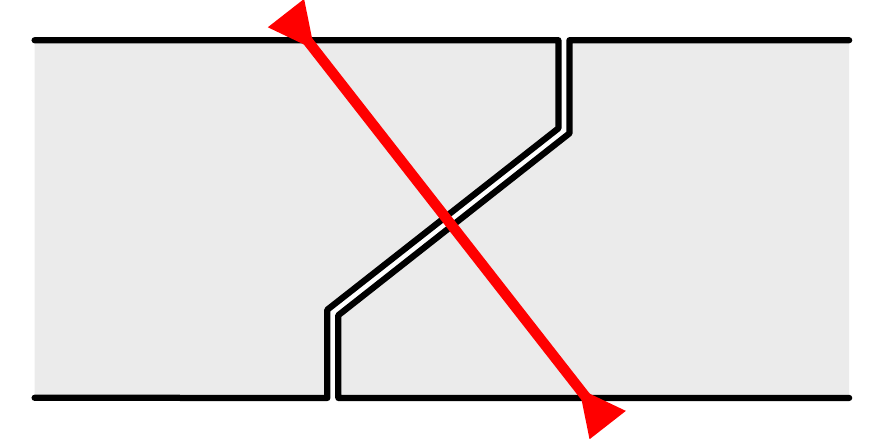}};
    \node[code] at ([xshift=0pt, yshift=-3pt]v3img.north west)
        {\begin{tabular}{@{}r@{}c*{5}{@{}c@{}c}@{}}
     & \textbf{J} & . & \textbf{O} & . & \textbf{I} & . & \textbf{N} & . & \textbf{T} & . & \textbf{S} \\
  3. & EE         & . & BB         & . & A          & . & 0S0        & . & BPB        & . & -
\end{tabular}};

    \node[anchor=north west, inner sep=0pt] (v4img)
        at ([xshift=4cm]v1img.north west)
        {\includegraphics[width=0.2\linewidth]{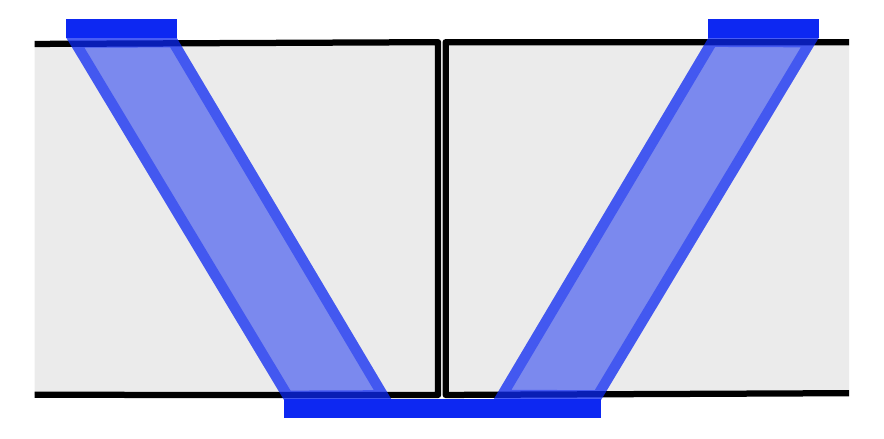}};
    \node[code] at ([xshift=0pt, yshift=-3pt]v4img.north west)
        {\begin{tabular}{@{}r@{}c*{5}{@{}c@{}c}@{}}
     & \textbf{J} & . & \textbf{O} & . & \textbf{I} & . & \textbf{N} & . & \textbf{T} & . & \textbf{S} \\
  4. & EE         & . & BB         & . & P          & . & 0S0        & . & CMC        & . & -
\end{tabular}};

    \node[anchor=north west, inner sep=0pt] (v5img)
        at ([xshift=4cm]v2img.north west)
        {\includegraphics[width=0.2\linewidth]{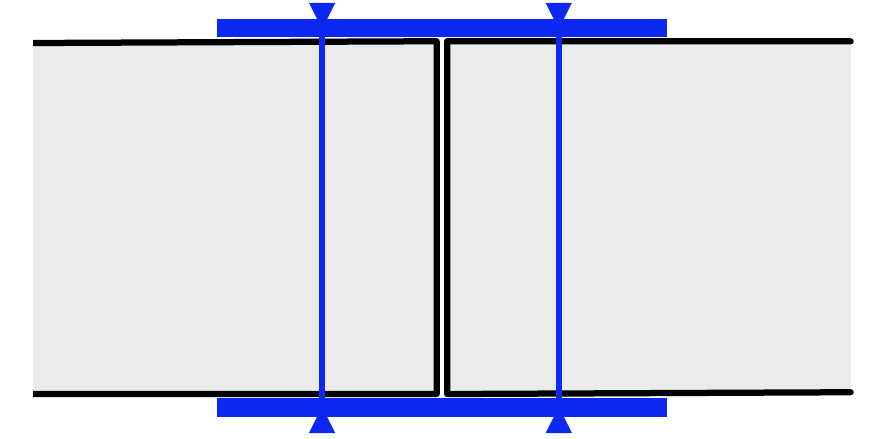}};
    \node[code] at ([xshift=0pt, yshift=-3pt]v5img.north west)
        {\begin{tabular}{@{}r@{}c*{5}{@{}c@{}c}@{}}
     & \textbf{J} & . & \textbf{O} & . & \textbf{I} & . & \textbf{N} & . & \textbf{T} & . & \textbf{S} \\
  5. & EE         & . & BB         & . & P          & . & 0S0        & . & BMB        & . & -
\end{tabular}};

    \node[anchor=north west, inner sep=0pt] (v6img)
        at ([xshift=4cm]v3img.north west)
        {\includegraphics[width=0.2\linewidth]{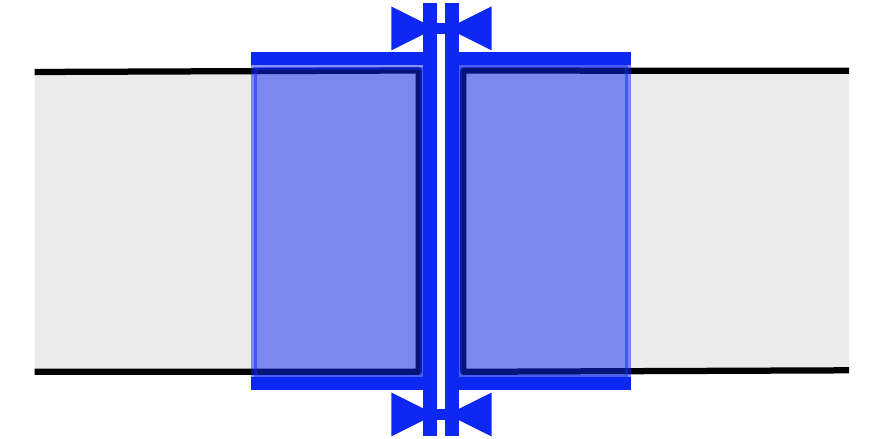}};
    \node[code] at ([xshift=0pt, yshift=-3pt]v6img.north west)
        {\begin{tabular}{@{}r@{}c*{5}{@{}c@{}c}@{}}
     & \textbf{J} & . & \textbf{O} & . & \textbf{I} & . & \textbf{N} & . & \textbf{T} & . & \textbf{S} \\
  6. & EE         & . & BB         & . & P          & . & 0S0        & . & FMF        & . & -
\end{tabular}};
\end{scope}

    \end{tikzpicture}
\caption{S: \textbf{S}core: Positioning of six representative connection variants in the
qualitative design space spanned by sustainability and
cost-efficiency.}
    \label{fig:variant-selection}
\end{figure}
Figure~\ref{fig:variant-selection} positions the six variants in the qualitative design space spanned by sustainability on the horizontal axis and cost-efficiency on the vertical axis. The interface determines whether the element ends can still be used in other juxtapositions and hence affects the connection sustainability. A planar end face can be joined to the side face of another element without further modification, which increases the reusability of such elements. The transfer mechanism determines whether the connecting member can be reused for elements with other dimensions and how much steel it requires. An enclosure follows the cross-section entirely and a post-tensioning member follows it only in its position. All connecting members are designed for the forces they transfer, but a post-tensioning member carries the force in pure tension, which tends to require less steel than members loaded in shear or bending. The positioning of each variant is justified below.

JOINT 1 (EE.BB.P.0S0.BPB) is a planar butt joint with a horizontal post-tensioning member bolted in recessed pockets at both element ends, adapted from \cite{BetonVerlag1987}. Its cost-efficiency is moderate. Manufacturing is limited to a drilled duct and a recessed pocket in each element. 
However, the assembly is difficult, because the space for tensioning in the recessed pocket is limited.
Its sustainability is high. The bolted member can be removed and reused separately from
the elements.

JOINT 2 (EE.BB.S.0S0.BPB) is a stepped butt joint with a vertical post-tensioning member bolted on the top and bottom faces, adapted from \cite{BetonVerlag1987}. Its cost-efficiency is lower than that of JOINT 1. In addition to the duct, two orthogonal cuts form the stepped interface at each element. No recessed pocket is needed as the member is tensioned on the top and bottom faces. 
The amount of added steel is similar. Its sustainability is also lower, because the stepped \textit{interface} requires a complementary counterpart in the next connection.

JOINT 3 (EE.BB.A.0S0.BPB) is an angled butt joint with an inclined post-tensioning member bolted on the top and bottom faces, adapted from \cite{Santner2016}. Its cost-efficiency is slightly lower than that of JOINT 2. The angled interface and duct require a higher manufacturing effort. In addition, the anchor plate needs an inclined recessed pocket or a wedge washer, whereas a flat anchor plate is sufficient for JOINT 2. 
Its sustainability is also slightly lower. The counterpart has to match the cut angle, and the inclined pocket is a further modification of the element.

JOINT 4 (EE.BB.P.0S0.CMC) is a planar butt joint with a steel enclosure that clamps around the element side faces, adapted from \cite{baghdadi2024}. Its cost-efficiency is lower than that of JOINT 1. 
The element ends need no modification, but
the enclosure requires considerably more steel than the post-tensioning member and has to be fitted tightly to the dimensions of the cross-section. Its sustainability is lower as well. The elements
remain unmodified and the enclosure can be removed, but the amount of added steel is large. The enclosure is also made for exactly one cross-section. If the counterpart in a reuse scenario has slightly
different dimensions, the enclosure cannot be attached.

JOINT 5 (EE.BB.P.0S0.BMB) is a planar butt joint with brackets bolted on the top and bottom faces, adapted from \cite{Kuepfer2022, Santner2016}. Its cost-efficiency is higher than that of JOINT 1. Manufacturing consists of a vertical drill hole in each element, and the brackets are bolted from accessible faces. The amount of added steel is higher, because the tension transferred through the connection loads the bolts in shear at the brackets. It is still considerably lower than that of the enclosure of JOINT 4. Its sustainability is slightly lower than that of JOINT 1 because of the amount of added steel, although the drill holes leave the end face unmodified and the brackets can be removed.

JOINT 6 (EE.BB.P.0S0.FMF) is a planar butt joint with a permanently fixed steel enclosure that joins both element ends through a bolted steel end plate, adapted from \cite{Kloft2024}. Its cost-efficiency is low. The enclosure requires slightly more steel than that of
JOINT 4 and has to be fitted and fixed to each element end. Its sustainability is low as well. The amount of added steel is large, and the permanent fixation means that element and enclosure cannot be reused separately. 
However, the end plate is highly standardized and can be adapted to other \textit{juxtapositions}.

In Figure~\ref{fig:variant-selection}, post-tensioning of JOINT 1 - 3 reaches moderate to high sustainability at moderate cost-efficiency.
The quantitative evaluation is therefore limited to the three variants with the \textit{transfer mechanism} \textbf{T}=BPB, i.e. JOINT 1, JOINT 2 and JOINT 3. Two further reasons lead to this selection.

First, post-tensioning places the \textit{interface} under compression before any external load acts. Its position and direction set where this compression acts in the cross-section and on which part of the interface. The compression can thus be placed where the external loading would otherwise open the \textit{interface}.

Second, post-tensioned dry joints are established in precast segmental construction. Analytical models for the anchorage of the prestressing force and for the \textit{interface} are available from this field and form the basis of the following mechanical evaluation.

\subsection{Procedure of mechanical evaluation}
\label{sec:procedure}

\begin{figure}[t!]
  \centering
  \includegraphics[width=1\linewidth]{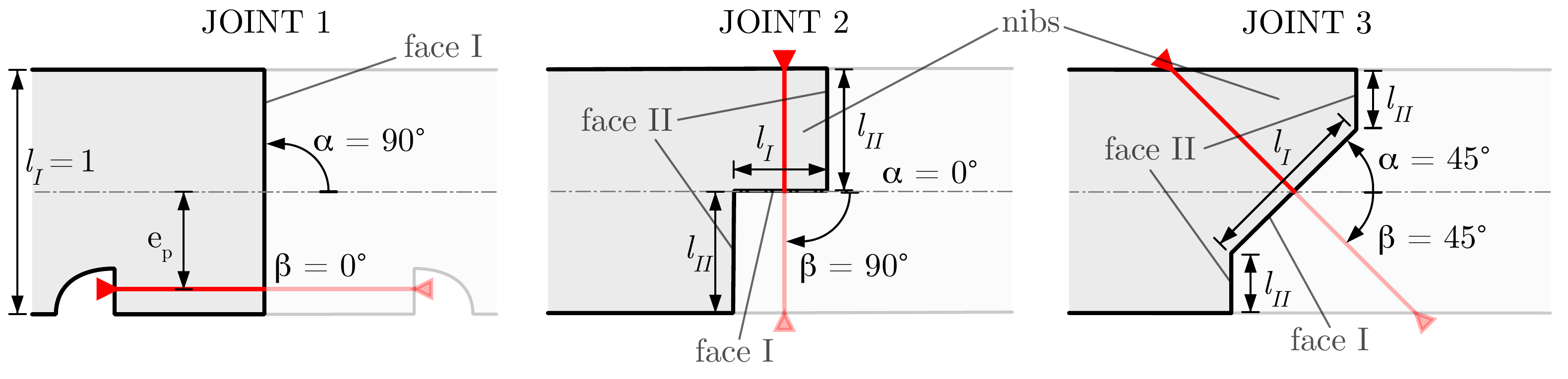}
\caption{Geometric parametrization of the three connection variants JOINT 1, JOINT 2 and JOINT~3 with the layout of the post-tensioning member, shown in red. Face~I of the interface has the length $l_\mathrm{I}$, face~II at JOINT 2 and JOINT 3 the length $l_\mathrm{II}$. All lengths are relative to the element height $h$.}
  \label{fig:param_dimensional}
\end{figure}
\begin{figure}[b!]
  \centering
  \includegraphics[width=0.9\linewidth]{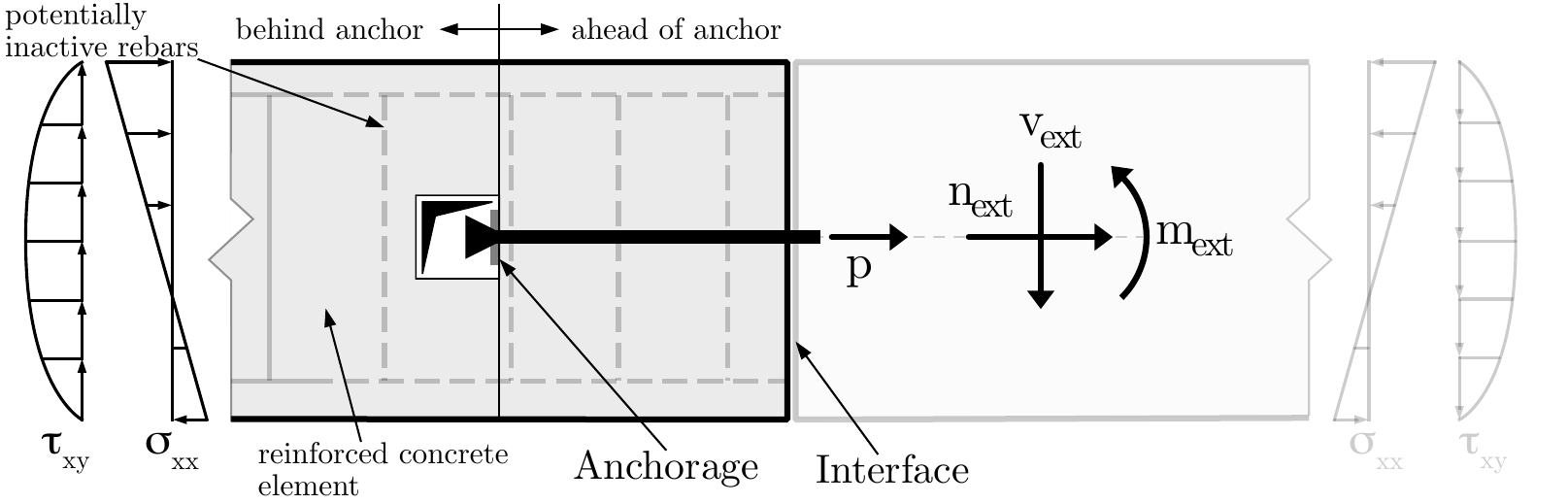}
  \caption{External section forces $n_\mathrm{ext}$,
$v_\mathrm{ext}$, $m_\mathrm{ext}$ acting on the left connection side and prestress $p$ in the tensioner of a connection of two
reused reinforced-concrete elements, with anchorage and interface. The far-field stresses are assumed to follow beam theory.}
  \label{fig:zones}
\end{figure}

 
For each variant, the evaluation determines the combinations of axial force, shear force, and bending
moment that the connection can transfer before a limit state is reached.
These are the external section forces $N_\mathrm{ext}$, $V_\mathrm{ext}$ and $M_\mathrm{ext}$ that have to be transferred through the connection. They are taken as known from the structural analysis. The prestressing force $P$ of the post-tensioning member is independent from the external section forces and enters
as a separate parameter. To make the result independent of the cross-section dimensions and the concrete strength, forces and moments are normalized as
\begin{equation}
  n_\mathrm{ext} = \frac{N_\mathrm{ext}}{f_\mathrm{c} \, b \, h}, \quad
  v_\mathrm{ext} = \frac{V_\mathrm{ext}}{f_\mathrm{c} \, b \, h}, \quad
  m_\mathrm{ext} = \frac{M_\mathrm{ext}}{f_\mathrm{c} \, b \, h^2}, \quad
  p = \frac{P}{f_\mathrm{c} \, b \, h},
  \label{eq:dimensionless_all}
\end{equation}
where $b$ and $h$ are the width and the height of the cross-section and $f_\mathrm{c}$ is the mean compressive strength of the reused reinforced-concrete element. The geometric parametrization of the three variants is shown in
Figure~\ref{fig:param_dimensional} and the positive
directions of the section forces are defined in Figure~\ref{fig:zones}.

The evaluation disregards the reinforcement inside the reused elements within the connection region as it may be unsuited for the present loading: The elements are reclaimed from the source building by cuts through a monolithic structure. Bars that cross a cut are exposed and lose their
anchorage at the cut face. Their anchorage is restored only by bonding
inside the element, away from the cut face. The connection that is assembled from these elements does not transfer stress through the exposed reinforcement bars. Stirrups close to
the cut can be preserved by a careful cut, but their condition after service, cutting, dismantling, and transport is not known. Every tensile component of the load transfer in the connection region is therefore assigned to the concrete. 

\begin{figure} [b]
    \centering
    \includegraphics[width=0.4\linewidth]{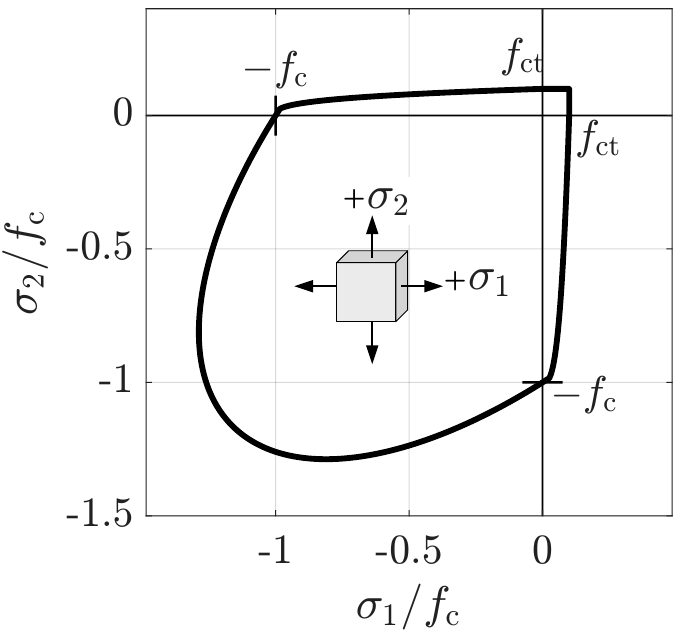}
  \caption{Biaxial failure envelope of \textcite{Kupfer1973}.}
  \label{fig:kupfer}
\end{figure}

The load capacity is the load at which a stress in the concrete first reaches
its strength. For tensile stresses this holds only where the exceedance of the tensile strength is detrimental to the load transfer.
Relying on the tensile strength of concrete for the load transfer for beam structures follows \textcite{Schlaich1987}, who utilize it where reinforcement cannot be provided. 
Where a tensile stress field in one direction is crossed by a compression
field in the other direction, they propose to consider the reduced biaxial strength. The computed stresses from the evaluation are therefore compared with the envelope of \textcite{Kupfer1973}, see Figure~\ref{fig:kupfer}. The tensile strength of the concrete is taken in proportion to its compressive strength as $f_\mathrm{ct} = 0.08\,f_\mathrm{c}$, which holds for the mean values of concrete C12/15 to C25/30 according to \cite{EN1992_1_1}.
The compression-compression region of the Kupfer envelope follows the elliptical branch
\begin{equation}
  \left(\frac{\sigma_1}{f_\mathrm{c}}\right)^{2}
  - 1.26\,\frac{\sigma_1}{f_\mathrm{c}}\,\frac{\sigma_2}{f_\mathrm{c}}
  + \left(\frac{\sigma_2}{f_\mathrm{c}}\right)^{2} = 1,
  \label{eq:kupfer-cc}
\end{equation}
and the compression-tension region follows the cube-root branch
\begin{equation}
  \frac{\sigma_1}{f_\mathrm{ct}}
  = \sqrt[3]{1 + \frac{\sigma_2}{f_\mathrm{c}}},
  \qquad \sigma_2 < 0,
  \label{eq:kupfer-ct}
\end{equation}
with the tensile strength under biaxial tension taken equal to
$f_\mathrm{ct}$.

\begin{table}[t!]
  \small
  \setlength{\tabcolsep}{3pt}
  \caption{Failure modes of anchorage and interface failure with their
  location, failure mechanism and references.}
  \label{tab:criteria}
\renewcommand{\tabularxcolumn}[1]{>{\raggedright\arraybackslash}m{#1}}
\begin{tabularx}{\linewidth}{@{}
    >{\raggedright\arraybackslash}m{0.18\linewidth}
    >{\raggedright\arraybackslash}m{0.14\linewidth}
    X
    >{\raggedright\arraybackslash}m{0.17\linewidth}
    >{\centering\arraybackslash}m{0.13\linewidth}@{}}
    \toprule
    Failure mode & Location & Failure mechanism & References & Sketch \\
    \midrule
    \multicolumn{5}{@{}l}{\textit{Anchorage failure (Section~\ref{sec:anchorage_failure})}} \\
    \addlinespace[-4pt]
    \ref{crit:bearing}~Bearing
      & beneath plate
      & crushing or splitting under concentrated compression
      & \textcite{Sanders1990}
      & \critfig{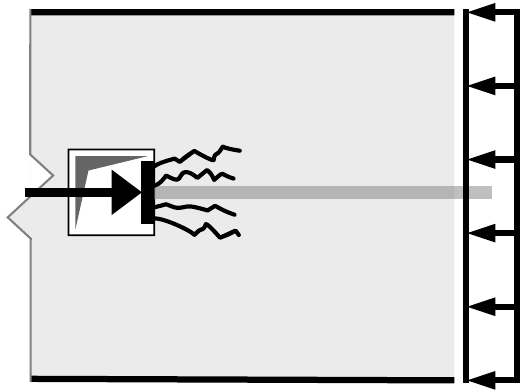} \\
    \ref{crit:bursting}~Bursting
      & on axis ahead of plate
      & splitting as force spreads into section
      & \textcite{Guyon1974}
      & \critfig{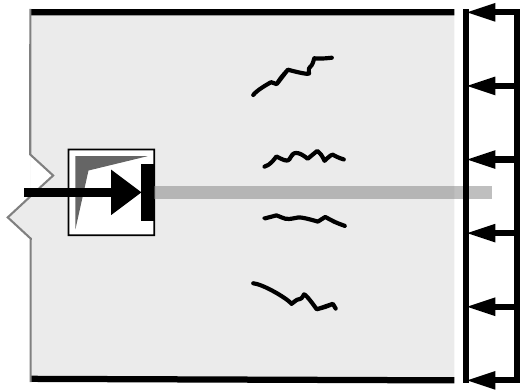} \\
    \ref{crit:spalling}~Spalling
      & free surface beside plate
      & crack at surface from inward displacement under plate
      & \textcite{Guyon1974}
      & \critfig{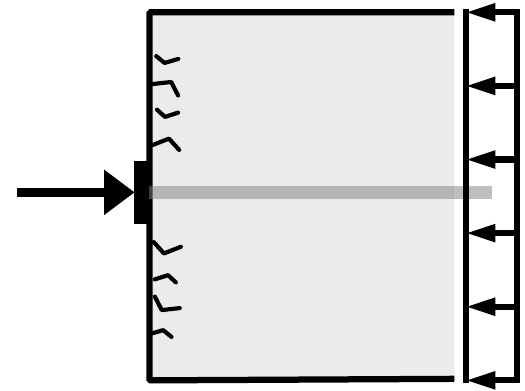} \\
    \ref{crit:tieback}~Tie-back
      & behind plate
      & crack as portion of anchorage force is carried back in tension
      & \textcite{Wollmann1992}
      & \critfig{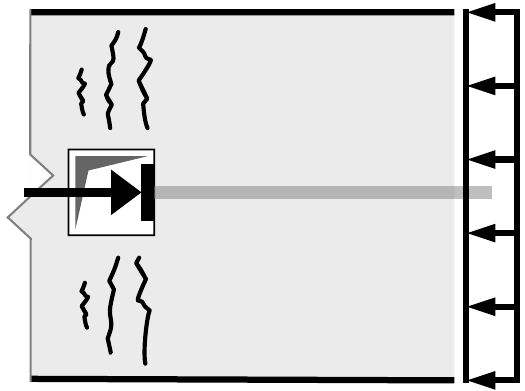} \\
    \ref{crit:edge}~Edge tension
      & edge remote from plate
      & crack under eccentric anchorage force
      & \textcite{Breen1994}
      & \critfig{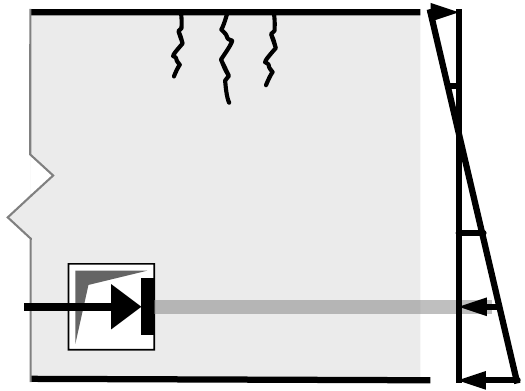} \\
    \midrule
    \multicolumn{5}{@{}l}{\textit{Interface failure, contact limit (Section~\ref{sec:contact})}} \\
    \addlinespace[-4pt]
    \ref{crit:sliding}~Sliding
      & faces I and II
      & sliding as traction component exceeds friction
      & Coulomb friction; \textcite{Santner2016}
      & \critfig{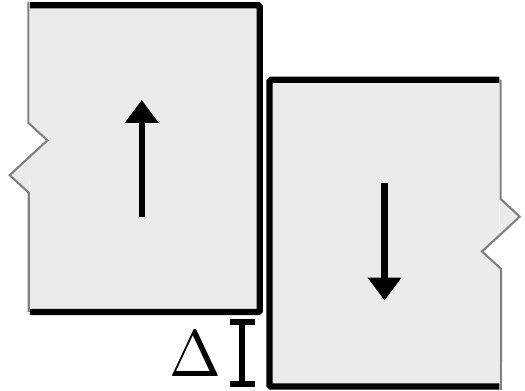} \\
    \ref{crit:decompression}~Decompression
      & face I
      & decompression from edge where contact stress would reach tension
      & linear contact stress; \textcite{Leonhardt1965}
      & \critfig{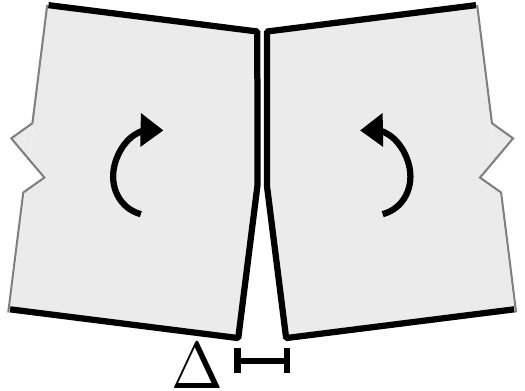} \\
    \midrule
    \multicolumn{5}{@{}l}{\textit{Interface failure, material limit (Section~\ref{sec:material_limit})}} \\
    \addlinespace[-4pt]
    \ref{crit:faces}~Face bearing
      & faces I and II
      & crushing or cracking under contact stresses
      & stress block; DAUB \cite{DAUB2024}
      & \critfig{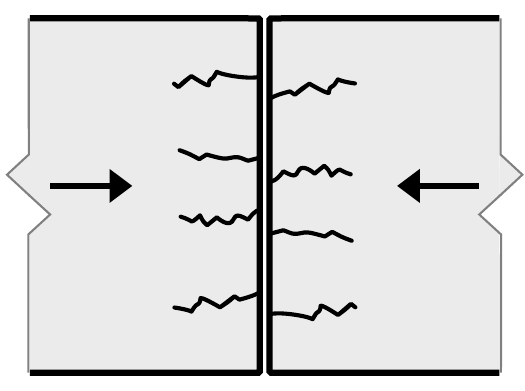} \\
    \ref{crit:root}~Interlocking
      & root of nib JOINT 2 and 3
      & shearing off or cracking under forces of interface
      & failure plane; \textcite{Rombach2002}
      & \critfig{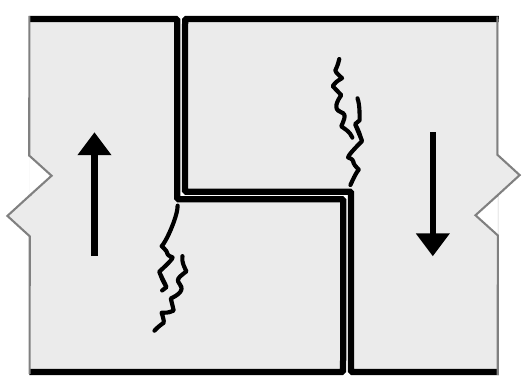} \\
    \bottomrule
  \end{tabularx}
\end{table}

The connection transfers the section forces through two mechanisms: The contact at the interface transmits a normal component in compression and a shear component by friction. The post-tensioning member acts in tension and transfers its force into the element through the anchorage. A high prestress as in post-tensioned bridge segments
is not treated here, since the anchorage is evaluated without reinforcement. Since the post-tensioning member is unbonded and anchored at
both ends, its force changes only if the distance between its
anchoring changes. As long as the interface remains in contact at the tendon axis and the slip along the interface remains small, this distance is unchanged, and the force remains at $p$. The evaluation therefore permits decompression of the interface up to the tendon axis only. Once the gap passes the tendon axis, further
rotation of the elements or a separation of the interface under tension elongates the post-tensioning member and raises its force above $p$. This additional force would have to be anchored, which the current evaluation of the anchorage failure does not cover.

This assumption separates the evaluation into two failure mechanisms -- anchorage failure
(Section~\ref{sec:anchorage_failure}) and interface failure
(Section~\ref{sec:interface_failure}). The anchoring is verified for the
prestressing force alone, which yields the range of $p$ it admits. The interface is then evaluated for interactions of $n_\mathrm{ext}$, $v_\mathrm{ext}$ and
$m_\mathrm{ext}$ at a fixed value of $p$. 

Table~\ref{tab:criteria} lists the failure modes of the interface and anchorage failure by location, mechanism and references.
The criteria of the first failure mechanism -- anchorage failure -- follow the transfer of the prestressing force from the anchorage plate into the element. The concrete beneath the anchorage plate
can fail from the concentrated compression, \ref{crit:bearing}. From the anchorage plate the force spreads over the section, which can then fail in transverse tension on the anchor axis, \ref{crit:bursting}. When the anchorage plate bears on a free surface, the concrete beneath the plate
is pressed into the element and the unloaded surface beside the plate must follow this inward displacement. This compatibility produces transverse tensile
stresses within a shallow zone at the surface, leading to cracks and possible failure \ref{crit:spalling}. If the element continues behind the anchorage plate, a portion of the
anchorage force is carried back in longitudinal tension into the concrete behind the plate, and may cause tie-back failure
\ref{crit:tieback}. When the anchorage plate is positioned eccentrically in the section, the edge
remote from the plate receives longitudinal tension, and may fail \ref{crit:edge}. 

The criteria of the second failure mechanism -- interface failure -- follow the force transferred from one element to the other. The interface transfers the traction component by friction, which can lead to sliding failure \ref{crit:sliding}. The decompression and contact opening of the interface makes the connection behave like a hinge that can lead to excessive deformations and eventual failure, \ref{crit:decompression}. The concrete adjacent to the
interface must carry the contact stresses, which can lead to compressive failure \ref{crit:faces}. This can be evaluated with the stress
block method as in the verification of the joints of tunnel segments \cite{DAUB2024}. For JOINT 2 and 3, the nibs interlock the elements and the forces of the interface pass through its root section, which can then fail \ref{crit:root}.

\subsection{Anchorage failure}
\label{sec:anchorage_failure}
The post-tensioning member is anchored through an anchorage plate at both element ends. The stresses it produces depend on the size of the plate relative to the section,
on the position of the plate and on its eccentricity. The three variants place the anchorage in three positions. For JOINT 1 the plate is eccentrically embedded in the
continuous element and reached through a recessed pocket. For JOINT 2 the plate is positioned on the top and bottom faces of the elements.  For JOINT 3, the inclined post-tensioning member is either clamped to an angled wedge on the top and bottom faces of the elements or its anchorage is installed inside the element and is accessible from the top and bottom side via a recess pocket, which is considered here. The anchorage plate is assumed square with side $a$ and introduces the force of a
single post-tensioning member into the element. The stresses that this force
induces in the element, (a) to (e) in Table~\ref{tab:criteria}, are analyzed in the following. \textcite{Breen1994} distinguish tensile stresses that are required for the
overall equilibrium of the anchorage zone from compatibility-induced stresses, which are not required for equilibrium and may be released by cracking. An induced stress bounds the force of the post-tensioning member only if its exceedance is detrimental to the strength of the anchorage.

\begin{figure} [b]
    \centering
    \includegraphics[width=0.4\linewidth]{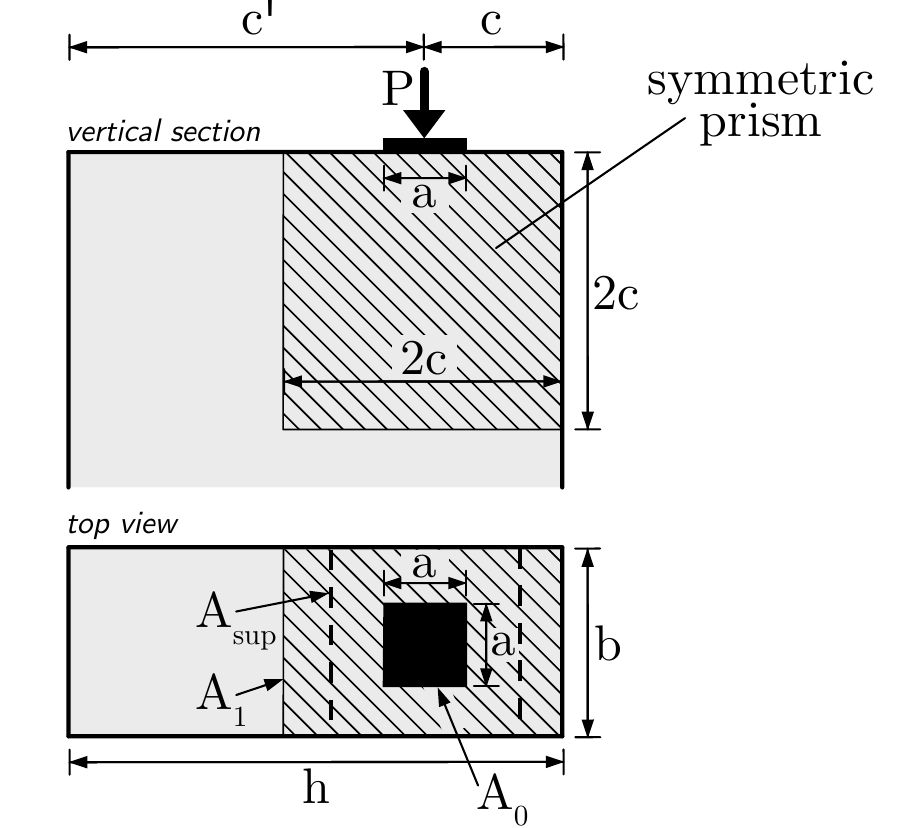}
\caption{Symmetric prism of the eccentric anchorage, adapted from
\textcite{Guyon1974}, with the net area $A_0$ of the bearing plate and its concentric supporting area $A_{\mathrm{sup}}$ inside the end block.}
  \label{fig:symmetric_prism}
\end{figure}

\crit{crit:bearing} Under partial-area loading, the unreinforced concrete fails in bearing in the small compression zone beneath the anchorage plate. There the triaxial stresses surpass the uniaxial strength of the concrete, as the surrounding concrete confines this zone. The average pressure under the plate is bounded by
\begin{equation}
  p_0 = \frac{P}{A_0} \le 0.8\,f_\mathrm{c} \sqrt{\frac{A_{\mathrm{sup}}}{A_0}}
  \le 3.0\,f_\mathrm{c},
  \label{eq:bearing}
\end{equation}
with $A_0$ the net area of the anchor plate and $A_{\mathrm{sup}}$ its concentric supporting area on the block, see Figure~\ref{fig:symmetric_prism}. $A_{\mathrm{sup}}$ is bounded by the smallest distance from the anchor axis to the block boundary in either direction.
According to experiments by Sanders~\cite{Sanders1990}, the coefficient 0.8 in Eq.~\eqref{eq:bearing} is the best fit to the results, although 0.7 is recommended as the design value. The confinement of the surrounding concrete raises the admissible pressure up to the bound $3.0\,f_c$.

\begin{figure}[t!]
  \centering
  \begin{subfigure}[b]{0.345\textwidth}
    \centering
    \includegraphics[width=\linewidth]{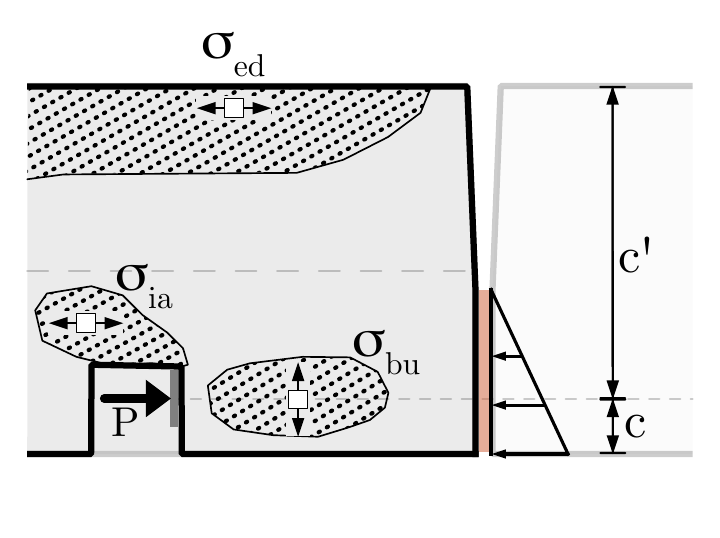}
    \caption{JOINT 1, vertical plane interface}
    \label{fig:anchorage_c1}
  \end{subfigure}
  \hfill
  \begin{subfigure}[b]{0.28\textwidth}
    \centering
    \includegraphics[width=\linewidth]{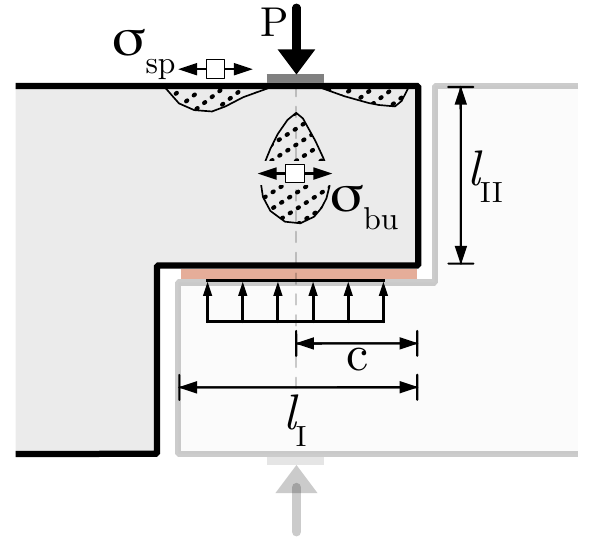}
    \caption{JOINT 2, stepped interface}
    \label{fig:anchorage_c2_1}
  \end{subfigure}
  \hfill
  \begin{subfigure}[b]{0.345\textwidth}
    \centering
    \includegraphics[width=\linewidth]{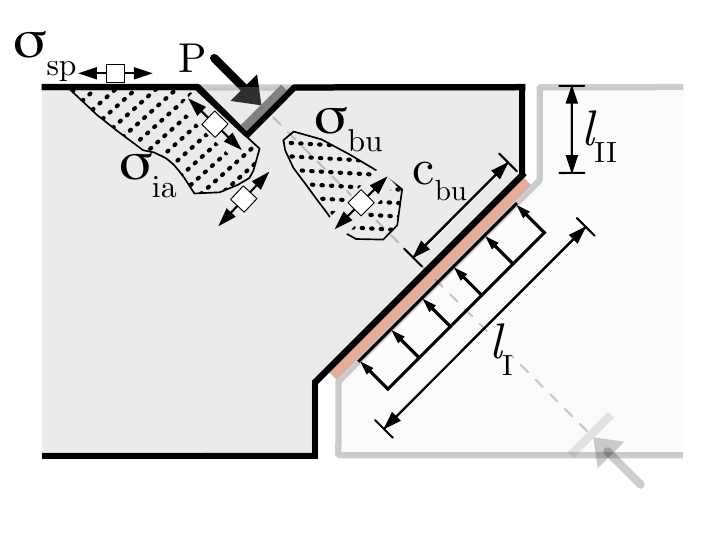}
    \caption{JOINT 3, angled interface}
    \label{fig:anchorage_c3_1}
  \end{subfigure}
  \caption{Locations of extremal stresses of the anchorage of JOINTS 1, 2 and 3.}
  \label{fig:anchorage_variants}
\end{figure}

\crit{crit:bursting} The spreading of the localized anchor force into the concrete member induces tensile bursting stresses ahead of the anchor, which may lead to
tensile splitting of the concrete. These stresses act normal to the post-tensioning member axis. Once the concrete splits along the post-tensioning member, the anchorage fails. Following Guyon~\cite{Guyon1974}, the elastic maximum of the bursting stress is
\begin{equation}
  \sigma_\mathrm{bu} = 0.517\, p_1 \left(1-\frac{a}{2c}\right)<f_\mathrm{ct},
  \qquad
  p_1 = \frac{P}{A_1}.
  \label{eq:sigma_bu}
\end{equation}
The solution confines the spreading of the force to a symmetric prism, see Figure~\ref{fig:symmetric_prism}. The prism is placed symmetrically under the post-tensioning member and has the depth $2c$, with
$c$ being the smaller distance of this axis from the block boundary. $A_1$ is the cross-section of the prism and $p_1$ the mean pressure on it.
Eq.~\eqref{eq:sigma_bu} holds for rectangular sections and is applied in both directions of the section, height and width, with $c$
taken in the respective direction, see Figure~\ref{fig:anchorage_variants}.
Where the element continues behind the anchor, as it is the case for JOINT 1 and JOINT 3, the maximum bursting stress at this intermediate anchorage is about 30\,\% lower than at the end anchorage, as for JOINT 2 ~\cite{Wollmann1992}. The reduced stress relies on the tie-back of the concrete volume behind the anchor and is not utilized in the evaluation.

\begin{figure}[b!]
  \centering
  \begin{subfigure}[t]{0.32\linewidth}
    \centering
    \includegraphics[width=\linewidth]{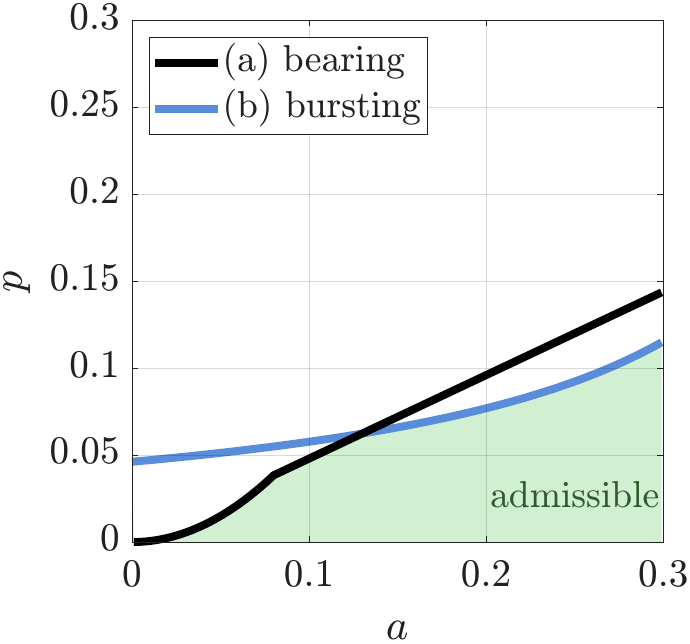}
    \caption{JOINT 1, $e_\mathrm{p} = 0.35$, $l_\mathrm{I} = 1$.}
    \label{fig:anchorage_window_C1}
  \end{subfigure}\hfill
  \begin{subfigure}[t]{0.32\linewidth}
    \centering
    \includegraphics[width=\linewidth]{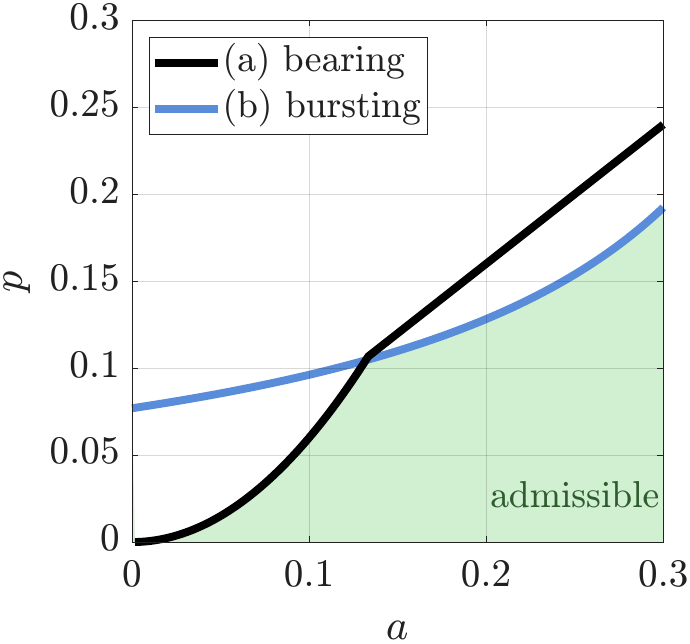}
    \caption{JOINT 2, $l_\mathrm{I} = 0.75$, $l_\mathrm{II} = 0.5$.}
    \label{fig:anchorage_window_C2}
  \end{subfigure}\hfill
  \begin{subfigure}[t]{0.32\linewidth}
    \centering
    \includegraphics[width=\linewidth]{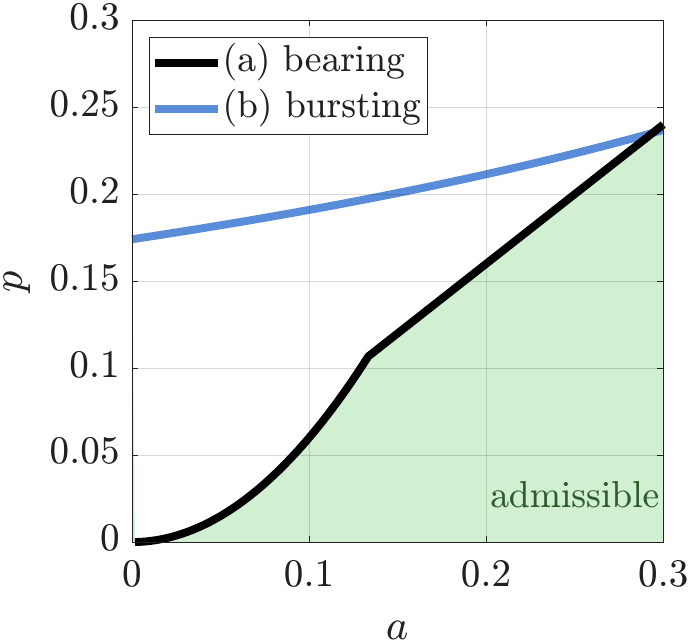}
    \caption{JOINT 3, $l_\mathrm{I} = 1.13$, $l_\mathrm{II} = 0.1$.}
    \label{fig:anchorage_window_C3}
  \end{subfigure}
  \caption{Anchorage failure: Results of the admissible prestress $p$ in dependence of the anchor plate size according to failure modes (a) and (b) for JOINTS 1, 2 and 3.}
  \label{fig:anchorage_windows}
\end{figure}

\crit{crit:spalling} At the end anchorage, as for JOINT 2 and in part for JOINT 3, the
inward displacement of the concrete beneath the plate induces transverse tension $\sigma_{\mathrm{sp}}$ along the loaded face adjacent to the plate,
see Figure~\ref{fig:anchorage_variants} and Appendix~\ref{app:spalling}. In principle, these spalling stresses are required to remain below $f_\mathrm{ct}$. They are, however, compatibility-induced
and not required for the overall equilibrium of the anchorage ~\cite{Breen1994}. The spalling stresses may be relieved by the formation of local cracking,
which leaves the spreading of the induced force undisturbed.

\crit{crit:tieback} At the intermediate anchorage, as for JOINT 1 and in part for JOINT 3, the concrete behind the plate transfers a portion of the induced force back in longitudinal tension $\sigma_{\mathrm{ia}}$ \cite{Wollmann1992}, see Figure~\ref{fig:anchorage_variants} and Appendix~\ref{app:tieback}. The tie-back tension $\sigma_{\mathrm{ia}}$ should remain below
$f_\mathrm{ct}$. Exceeding it is not detrimental to the anchorage if the
anchorage is also verified for the higher bursting stress of the end
anchorage. Where the tie-back tension exceeds the tensile strength of the concrete, the portion tied back passes ahead of the anchorage through the compression path, which remains
available~\cite{Wollmann1992}. The bursting stress of the end anchorage position in (b) is therefore evaluated to all three connection variants.

\crit{crit:edge} Where the anchor is positioned eccentrically, as for JOINT 1, tensile
stresses $\sigma_{\mathrm{ed}}$ act parallel and close to the longitudinal
edge remote from the anchor plate~\cite{Breen1994}, see Appendix~\ref{app:edge}. This longitudinal edge tension is required for equilibrium of the section but lies outside the path of the anchor force. Where it exceeds the tensile strength of the concrete $f_\mathrm{ct}$, the section reduces to its compressed depth. The edge closes again under the external loads that return it to compression.

The exceedance of the tensile strength by the stresses of (c) to (e) is not detrimental to the transfer of the anchor force. The admissible prestress therefore follows from criteria
(a) and (b) alone. Figure~\ref{fig:anchorage_windows} shows the evaluation of criteria (a) and (b) for the square plate with size $a$ and a section width $b = 0.5$, considering the
three anchorage positions of JOINT 1, JOINT 2 and JOINT 3. For JOINT 1 the
bearing failure bounds the prestress up to $a \approx 0.12$, bursting failure beyond, with about $p = 0.11$ at
$a = 0.3$. For JOINT 2 the symmetric prism is assumed within the nib, $2c = l_{II}$. Bursting failure takes over at $a \approx 0.13$ and admits about $p = 0.19$ at
$a = 0.3$. For JOINT 3 the symmetric prism is placed on the inclined face~I, $2c = l_I$. Bearing failure bounds the prestress up to $a \approx 0.29$. The interface failure limit for all connection variants can therefore be evaluated for a prestress level of up to $p = 0.10$.
\subsection{Interface failure}
\label{sec:interface_failure}

A dry interface between concrete faces transmits compressive contact stress and friction. Interface failure can accordingly occur in two distinct ways. The contact limit (Sec.~\ref{sec:contact}) is reached when a face
decompresses or slides and no other face is available for the transfer of forces. The material limit (Sec.~\ref{sec:material_limit}) is reached when the contact
stresses that equilibrium demands exceed the strength of the concrete adjacent to the interface.

Figure~\ref{fig:param_dimensional} shows the faces of the interface.
Which faces are in contact is a question of fabrication accuracy. Match-cast or ground joints of segmental bridges reach their fit of 0.1\,mm only through
fine processing of the faces~\cite{Santner2016}. Reuse elements sawn out of existing structures do not possess such geometric interface accuracy. Form deviations of the interface on the order of the cutting tolerance must
be expected~\cite{Schaarschmidt2023}. Of these only the length deviation of the nibs of JOINT 2 and 3 between the two element ends is
considered here. It directly determines which faces are in contact. 
Deviations of the cutting angles and of the remaining dimensions are disregarded. Two contact states are distinguished therefore in this work. 
State 1 is defined by the contact that the prestress establishes at
installation. State 2 is defined by the contact after the external forces have brought a further part of the interface into contact. For JOINT 1 the
eccentric post-tensioning member compresses only a part of face I in state 1. State 2 begins once the external forces compress the entire face I. For JOINT 2 and 3 the post-tensioning member presses face I into contact
and a gap remains at face II. Only face I is therefore in contact in state 1.
State 2 begins once the element slides along face I and closes the gap at the nearer part of face II.

\subsubsection{Contact limit}
\label{sec:contact}

\begin{figure}[b!]
  \centering
\includegraphics[width=\linewidth]{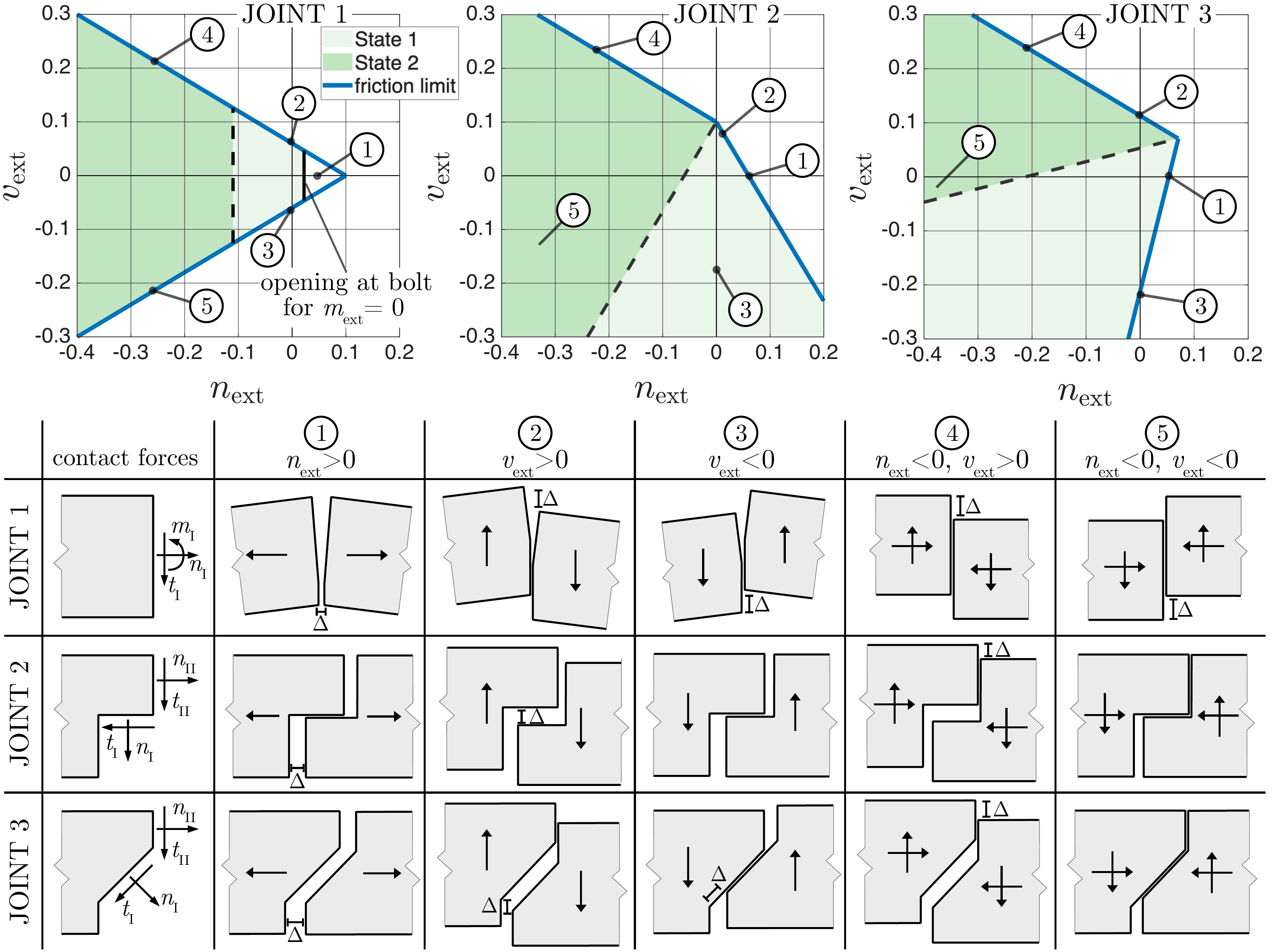}
\caption{Interface failure: Contact limit from criteria (f) and (g) of Table~\ref{tab:criteria} for interactions of $n_\mathrm{ext}$ and $v_\mathrm{ext}$ at
  $m_\mathrm{ext} = 0$, shown for prestress $p = 0.1$. The table indicates the contact forces of
  faces I and II with their positive directions and the relative displacement $\Delta$ of the contact faces for load cases \textcircled{\scriptsize 1} to \textcircled{\scriptsize 5}.}
  \label{fig:nv_interaction}
\end{figure}

The contact conditions are formulated for the force transferred across the joint. Its contact forces $n_I$, $t_I$ on face~I and $n_{II}$, $t_{II}$ on
face~II are defined in Figure~\ref{fig:nv_interaction}. In state 1, decomposing the external forces and the prestress on face~I and taking moments about its
centroid on the element axis gives
\begin{align}
  n_I &= n_\mathrm{ext} \sin\alpha + v_\mathrm{ext} \cos\alpha - p \sin(\alpha + \beta),
  \label{eq:nI}\\
  t_I &= - n_\mathrm{ext} \cos\alpha + v_\mathrm{ext} \sin\alpha + p \cos(\alpha + \beta),
  \label{eq:tI}\\
  m_I &= m_\mathrm{ext} - p\, e_p ,
  \label{eq:mI}
\end{align}
with $0^\circ \le \alpha \le 90^\circ$ and $0^\circ \le \beta \le 90^\circ$.
In state 2 face~I remains at its frictional capacity. Face~II bears the
share of the transferred force that exceeds this capacity. Face~I continues to bear its normal force. State 2 lasts as long as face~II bears the load. In both states each compressed face $j \in \{\mathrm{I}, \mathrm{II}\}$ is subject to two criteria, its
friction limit \ref{crit:sliding} and the compression that keeps it in contact \ref{crit:decompression}.

\crit{crit:sliding} On a dry face the Coulomb friction law without cohesion limits the
traction component to
\begin{equation}
  |t_j| \le -\mu\, n_j.
  \label{eq:sliding}
\end{equation}
Reported friction
coefficients $\mu$ of dry concrete joints differ widely, from $0.4$ to
$1.1$~\cite{Buyukozturk1990, Schaarschmidt2023}. Rombach's design model \cite{Rombach2002} for
keyed dry joints uses $0.65$. The subsequent evaluation adopts
$\mu = 0.6$ for all faces.

\crit{crit:decompression} A dry face transfers force only under compression, $n_j \le 0$. Its contact depth $x$ follows from the linear contact stress of the rigid element, as for the concrete hinges of \textcite{Leonhardt1965}. They allow the hinge to open to half the
cross-section height.
Here the face is allowed to open to the axis of the post-tensioning member. For JOINT 2 and 3 this axis lies at the middle of face~I, $l_I/2$ from either edge and for JOINT 1 it lies $e_p$ off the middle. The contact depth is therefore bounded by
\begin{equation}
  x \ge \frac{l_I}{2} \pm e_p.
  \label{eq:depth}
\end{equation}
The minus sign applies when the gap opens at the edge remote from the post-tensioning member. The depth follows from the eccentricity $e_I = |m_I/n_I|$ of the transferred force. Within the kern, $e_I \le l_I/6$, the whole face is under compression, $x = l_I$. Beyond it
the face decompresses from the edge to $x = 3\,(l_I/2 - e_I)$. 
The bound applies to face~I only. On which side of face~II the gap closes first
under an external moment is unknown in both states. The transfer of an external moment is therefore assigned only to face~I, whose contact is certain.

Figure~\ref{fig:nv_interaction} shows the evaluation of criteria (f) and (g) for interactions of $n_\mathrm{ext}$ and $v_\mathrm{ext}$ at $m_\mathrm{ext} = 0$ for
the selected connection variants. The influence of $m_\mathrm{ext} \neq 0$ on the contact limit is shown in Appendix~\ref{app:mext}.
The admissible region of JOINT 2 and JOINT 3 thus exceeds that of JOINT 1.
Whether the concrete material can transfer the forces of the admissible region is evaluated next.

\subsubsection{Material limit}
\label{sec:material_limit}

\begin{figure}[b]
  \centering
  \includegraphics[width=1\linewidth]{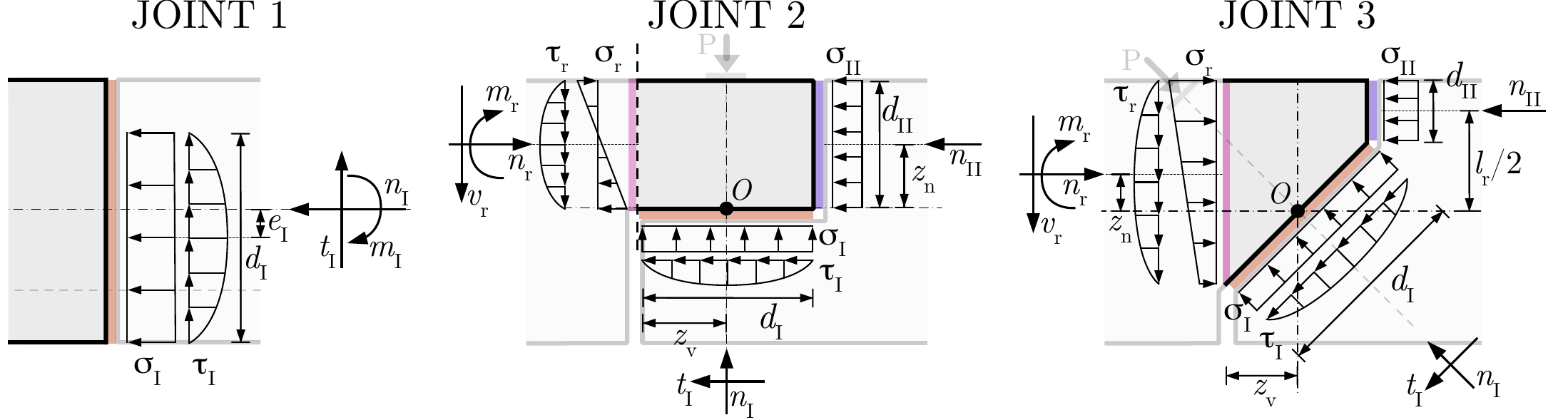}
  \caption{Assumed contact stresses and distributions at the root section for load
  direction \textcircled{\raisebox{-0.6pt}{\scriptsize 5}}. At JOINT 2 and JOINT 3, $O$ is the centre of face~I and the reference point for moment
    equilibrium in Eq.~\eqref{eq:root_moment}.}
  \label{fig:interface_stresses}
\end{figure}

The concrete transfers the contact forces of Sec.~\ref{sec:contact} at up to three locations, face~I, face~II and the root section, see Figure~\ref{fig:interface_stresses}. Face~I is present in all variants. Face~II and the root section are only present in JOINT~2 and 3, where the nib joins the full cross-section. The root sections corresponds to the failure plane at the base of shear keys in \textcite{Rombach2002}. At each location the forces
produce a normal stress $\sigma$ and a shear stress $\tau$. Both are normalized by concrete strength $f_\mathrm{c}$.

\crit{crit:faces} At the contact faces the stresses may reach the concrete strength. In state 1 the contact forces of face~I are those of
Eqs.~\eqref{eq:nI}--\eqref{eq:mI}. In state 2 the contact forces of both
faces follow from Appendix~\ref{app:contact}. The contact stresses are
evaluated with the stress block method used in the verification of the joints of tunnel segments (DAUB \cite{DAUB2024}). At high compression the stress-strain relation of the concrete may become
nonlinear. A constant stress block is therefore assumed in place of the linear contact
stress of (g). The normal component $n_j$ and the moment $m_j$ of a face are
represented by a constant stress block centred on their resultant. On face~I the
width $d_{\mathrm{I}}$ of the block decreases with the eccentricity $e_{\mathrm{I}}$ as $d_{\mathrm{I}} = l_{\mathrm{I}} - 2e_{\mathrm{I}}$.
On face~II the block width is $d_{\mathrm{II}} = l_{\mathrm{II}}$ since face~II transfers no moment. The
shear stress is conservatively taken parabolic over the same width. Its maximum is three halves of the mean. With the contact forces
$n_j$ and $t_j$ of a face this gives
\begin{equation}
  \sigma_j = \frac{n_j}{d_j}, \qquad
  \tau_j = \frac{3}{2}\frac{|t_j|}{d_j}, \qquad
  j \in \{\mathrm{I}, \mathrm{II}\}.
  \label{eq:face_stress}
\end{equation}

\begin{figure}[t!]
  \centering
  \begin{subfigure}[t]{0.325\linewidth}
    \centering
    \includegraphics[width=\linewidth]{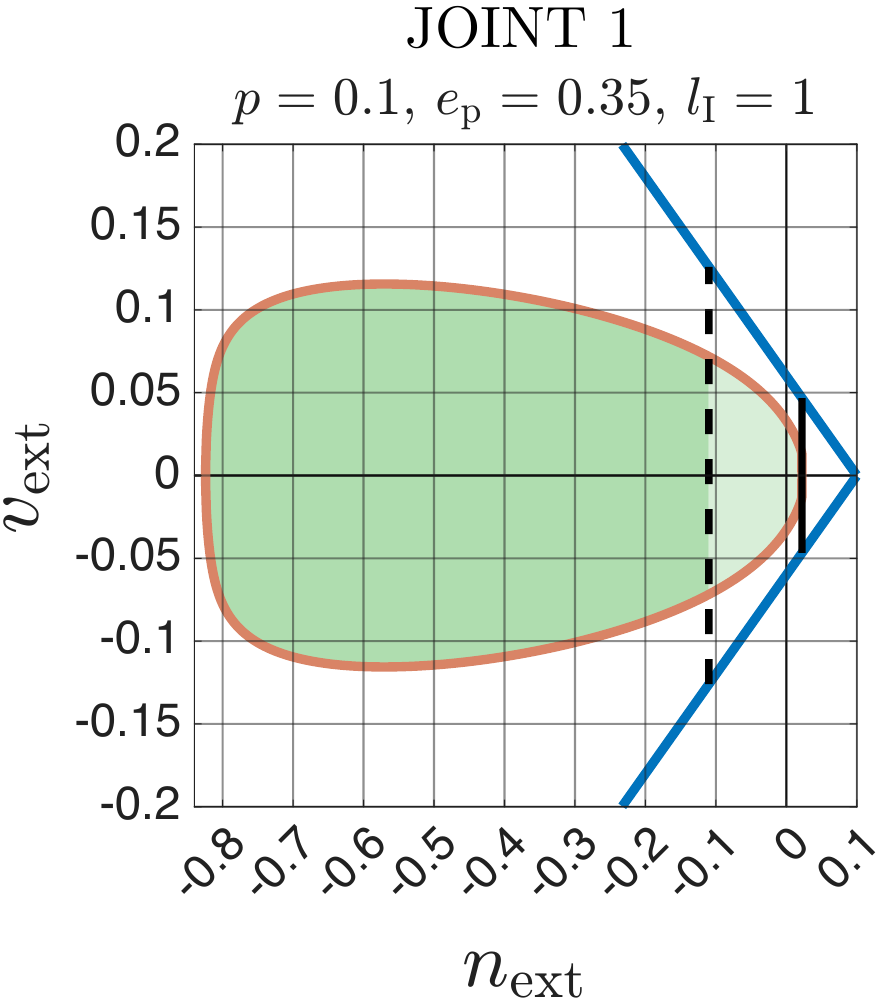}
    \label{fig:material_limit_C1}
  \end{subfigure}\hfill
  \begin{subfigure}[t]{0.325\linewidth}
    \centering
    \includegraphics[width=\linewidth]{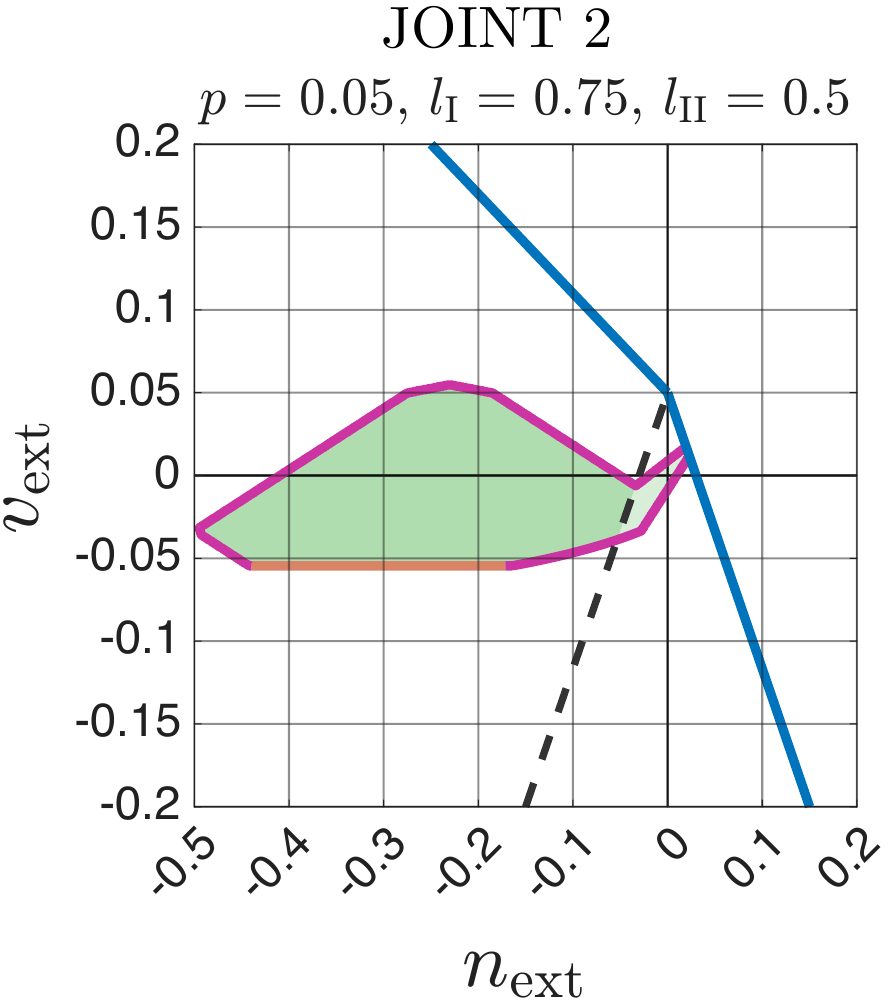}
    \label{fig:material_limit_C2}
  \end{subfigure}\hfill
  \begin{subfigure}[t]{0.325\linewidth}
    \centering
    \includegraphics[width=\linewidth]{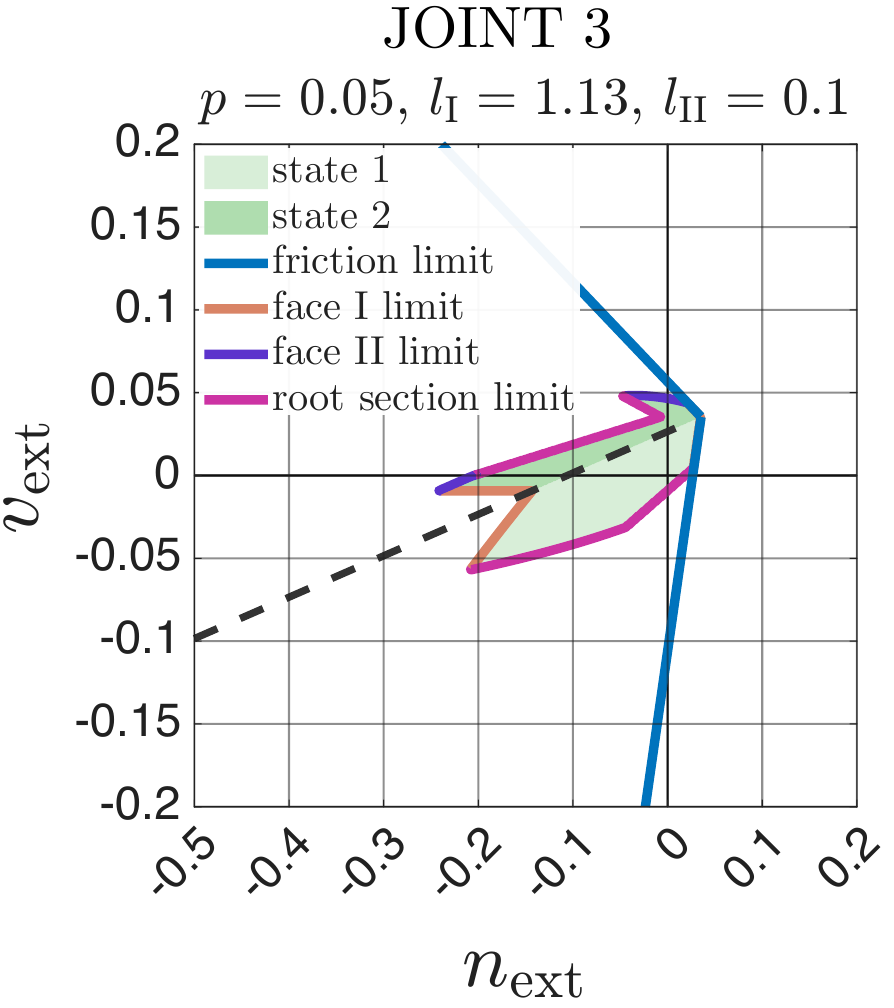}
    \label{fig:material_limit_C3}
  \end{subfigure}
  \caption{Exemplary evaluation of material limit for interaction of $n_\mathrm{ext}$ and $v_\mathrm{ext}$ with $m_\mathrm{ext}=0$.}
  \label{fig:material_limit}
\end{figure}
\crit{crit:root} For JOINT 2 and 3 the root section transfers the forces of
both faces. Its stresses may reach the concrete strength as well. Its section forces $n_r$ and $v_r$ follow from equilibrium of the nib. For
JOINT 2 the anchorage is positioned inside the nib and the
prestress is not transferred through the root section. The root section then transfers only the external forces $n_r = n_{\mathrm{ext}}$ and
$v_r = v_{\mathrm{ext}}$. At JOINT 3 the prestress is transferred
through the root section. Its section forces are
$n_r = n_{\mathrm{ext}} - p\cos\beta$ and
$v_r = v_{\mathrm{ext}} - p\sin\beta$. The moment at the root section follows from moment equilibrium of the nib about the reference point $O$ of face~I, Figure~\ref{fig:interface_stresses}, as
\begin{equation}
  m_r = m_{\mathrm{ext}} + n_r\, z_n
        - \left(v_r + t_{\mathrm{II}}\right) z_v
        - n_{\mathrm{II}}\, \frac{l_r}{2}.
  \label{eq:root_moment}
\end{equation}
Here, $z_n = l_{\mathrm{II}}/2$ and $z_v = l_{\mathrm{I}}\cos\alpha/2$ are the
lever arms of $n_r$ and $v_r$. Unlike the contact faces the root section lies in the monolithic concrete and can transfer tension. 
$\sigma_r$ is therefore taken linear and $\tau_r$ parabolic over $l_r$. The stresses are verified at their extremes,
\begin{equation}
  \begin{aligned}
    \sigma_r &= \frac{n_r}{l_r} \pm 6\frac{m_r}{l_r^2},
    & \tau_r &= 0 && \text{at the edges,}\\
    \sigma_r &= \frac{n_r}{l_r},
    & \tau_r &= \frac{3}{2}\frac{|v_r|}{l_r} && \text{at the centre.}
  \end{aligned}
  \label{eq:root_stress}
\end{equation}
At every location the normal and the shear stress are transformed into
principal stresses, see Appendix~\ref{app:verify}. These are compared with the biaxial failure envelope of
Kupfer \cite{Kupfer1973}. Figure~\ref{fig:material_limit} shows the result of the material limits for $m_{\mathrm{ext}} = 0$, evaluated with a prestress of
$p = 0.1$ for JOINT 1 and of $p = 0.05$ for
JOINT 2 and 3. The three variants differ in the transfer of the shear forces. 
For JOINT 1 the shear is transferred across the joint face over the full depth of the element. For JOINT 2 and 3 face~I bears the external shear force first. From there it passes through the root section into the rest of the element. The root section is shallower than the element, half its depth for
JOINT 2 and three quarters for JOINT 3. The same force is transferred through less concrete. The stresses rise accordingly. The
geometric interlock of the nibs of JOINT 2 and 3 raises the contact limit compared to JOINT 1. The material limit leaves a much smaller admissible region. The lower prestress of JOINT 2 is necessary. In state 1 face~I transfers $n_{\mathrm{ext}}$ by friction alone. This friction acts at the step face and increases the root moment $m_r$. For $p = 0.1$ the frictional capacity of face~I is higher.
A higher $n_{\mathrm{ext}}$ is then needed before the joint passes to state~2, where face~II comes into contact. The root section reaches the tensile strength first. Hence state 2 is not reached within the material limit.

\section{Conclusion}
\label{sec:discussion}
This paper proposes a systematic classification for detachable dry connections together with analytical results that identify and evaluate a variety of different failure mechanisms of post-tensioned connection variants.

The classification, denoted JOINTS and applicable to concrete connections beyond the reuse context, organizes the design space along five categories whose initials spell JOINT: the Juxtaposition as the spatial arrangement of the paired elements, the Origin as their prior structural function, the Interface geometry, the Newly introduced materials at the joint and the Transfer mechanism. A sixth category, the Score, completes the acronym and rates each configuration according to sustainability, cost-efficiency, and mechanical performance.

The mechanical performance of three selected connection variants has been evaluated analytically using engineering criteria.
Since the reused elements lack appropriate reinforcement near the cut surface, the concrete must sustain combined compression and shear at the connection region alone. A crack that disturbs the load path therefore disrupts
the force transfer. At the anchorage the load path is the compression spreading from the anchorage plate into the element section. It fails by
crushing under the plate or by the bursting tension along the post-tensioning member. The cracks of spalling, tie-back and edge tension lie outside the load path, so that the prestress can still be transferred. The admissible
prestress of compact plates is therefore bounded by bearing and that of larger plates by bursting.

At the interface the contact limit follows from the friction of the faces and the compression that keeps them in contact. Two contact states are
analysed to account for the form deviation of the nibs. The geometric interlock of the nibs of JOINT 2 and 3 raises
the contact limit compared to JOINT 1. The material limit leaves a much smaller admissible region. In the high-compression range the biaxial strength of the concrete governs before interlock and friction are fully mobilized.

Two limitations frame the results. First, the presented stress values and distributions of the anchorage and interface failure modes are based on
engineering criteria and only approximately match the real stress state of the connection. However, they still allow to indicate the location of failure and approximate load capacity.  Secondly, complex interface geometries such as curved or finely profiled faces lead to more complex stress distributions on the contact interface, and shape deviations from cutting exclude the assumption of full-area contact, so a pointwise stress prediction is not available analytically. Reused elements furthermore retain reinforcement designed for the loading of the source structure rather than for the new joint location. This reinforcement is not absent, as assumed here, but its contribution to the local force transfer cannot be quantified within the present framework. These points call for more detailed numerical studies in future work.

\section*{Acknowledgements}
This research is funded by the Deutsche Forschungsgemeinschaft (DFG, German Research Foundation) within the Collaborative Research Centre CRC~1683 ``Methods of interaction for the modular reuse of existing load-bearing structures'' -- project number 523205694 -- which is gratefully acknowledged.
\appendix
\renewcommand{\thesection}{Appendix~\Alph{section}}
\renewcommand{\thesubsection}{\Alph{section}.\arabic{subsection}}
\renewcommand{\theequation}{\Alph{section}.\arabic{equation}}
\renewcommand{\thefigure}{\Alph{section}.\arabic{figure}}
\renewcommand{\thetable}{\Alph{section}.\arabic{table}}
\setcounter{equation}{0}
\setcounter{figure}{0}
\setcounter{table}{0}

\section{Failure modes (c) to (e) of the anchorage}
\label{app:anchorage}

This appendix considers the tensile stresses of (c) to (e) of Sec.~\ref{sec:anchorage_failure}. They do not bound the admissible prestress.
They indicate the prestress at which the concrete cracks outside the load path at the installation state.

\subsection{Spalling (c)}
\label{app:spalling}
Following \textcite{Guyon1974}, the spalling stress resultant is
\begin{equation}
  T_\mathrm{sp} = \left[0.04 + 0.20
  \left(\frac{c'-c}{c'+c}\right)^{3}\right] P ,
  \label{eq:T_sp}
\end{equation}
with $c'$ being the larger distance of the anchor axis from the block boundary, see Figure~\ref{fig:symmetric_prism}. For an anchorage plate spanning the width of the element $b$, the maximum spalling stress is
\begin{equation}
  \sigma_\mathrm{sp} = \frac{2\,T_\mathrm{sp}}{0.1\,(c+c')\,b} < f_\mathrm{ct}\,.
  \label{eq:sigma_sp}
\end{equation}
For an anchorage plate of a the compact size, which is not covering the full element width, the tension beside the plate is a near-field quantity. According to \textcite{Timoshenko1951}, the free surface
outside a load $P$ on a half-space carries the radial tension $(1-2\nu)P/(2\pi r^{2})$, which at the edge of the circular plate of
equal area, $r^{2} = a^{2}/\pi$, becomes $\sigma_\mathrm{sp} = (1-2\nu)\,p_0/2<f_\mathrm{ct}$. The radial stress is evaluated in addition to Eq.~\eqref{eq:sigma_sp}. The
larger of the two stresses is verified against the tensile strength.

\begin{figure}[b!]
  \centering
  \begin{subfigure}[t]{0.58\linewidth}
    \centering
    \includegraphics[width=\linewidth]{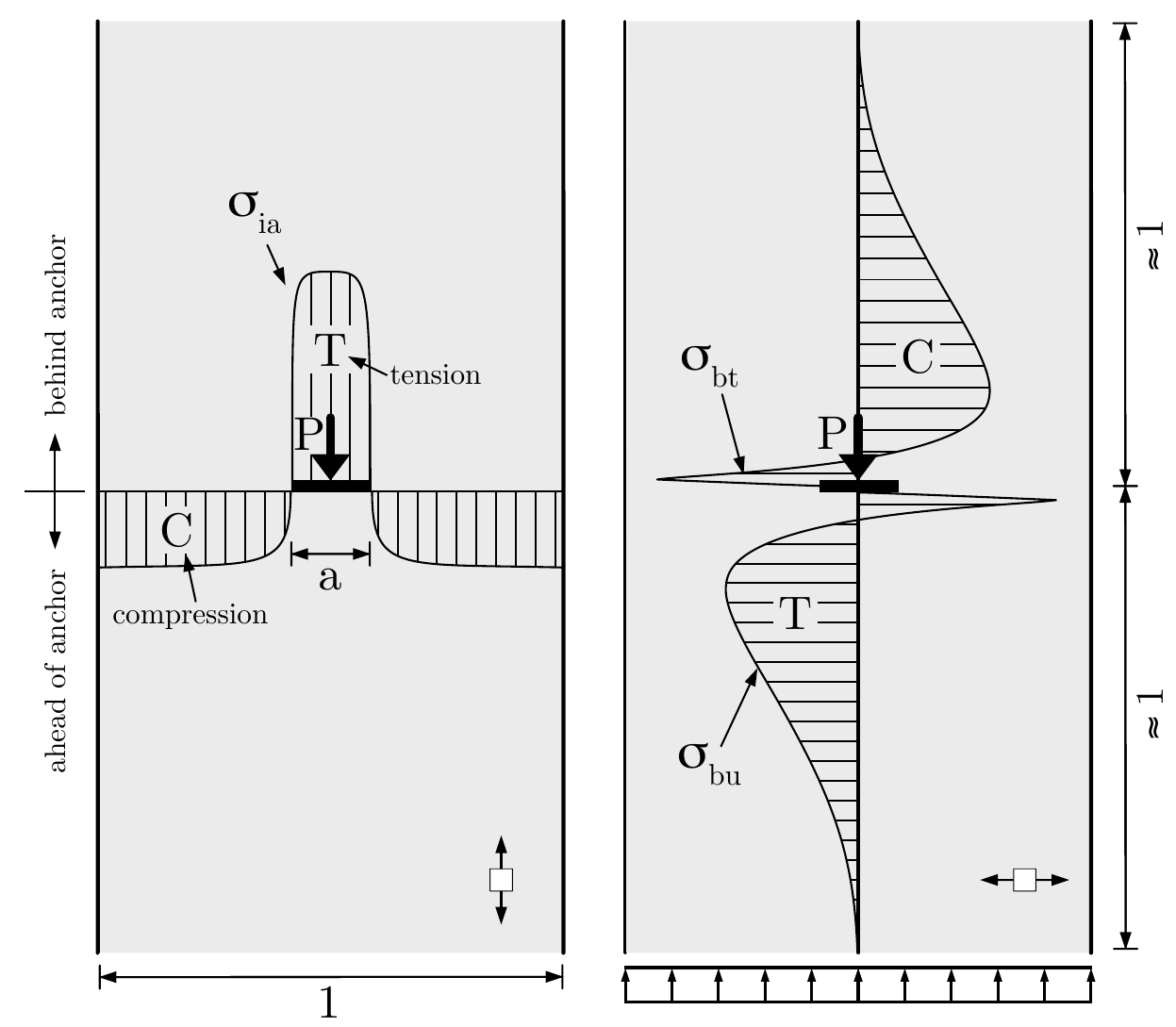}
    \caption{Longitudinal and transverse stresses at an intermediate anchorage.}
    \label{fig:stress_state_intermediate}
  \end{subfigure}
  \hfill
  \begin{subfigure}[t]{0.35\linewidth}
    \includegraphics[width=\linewidth]{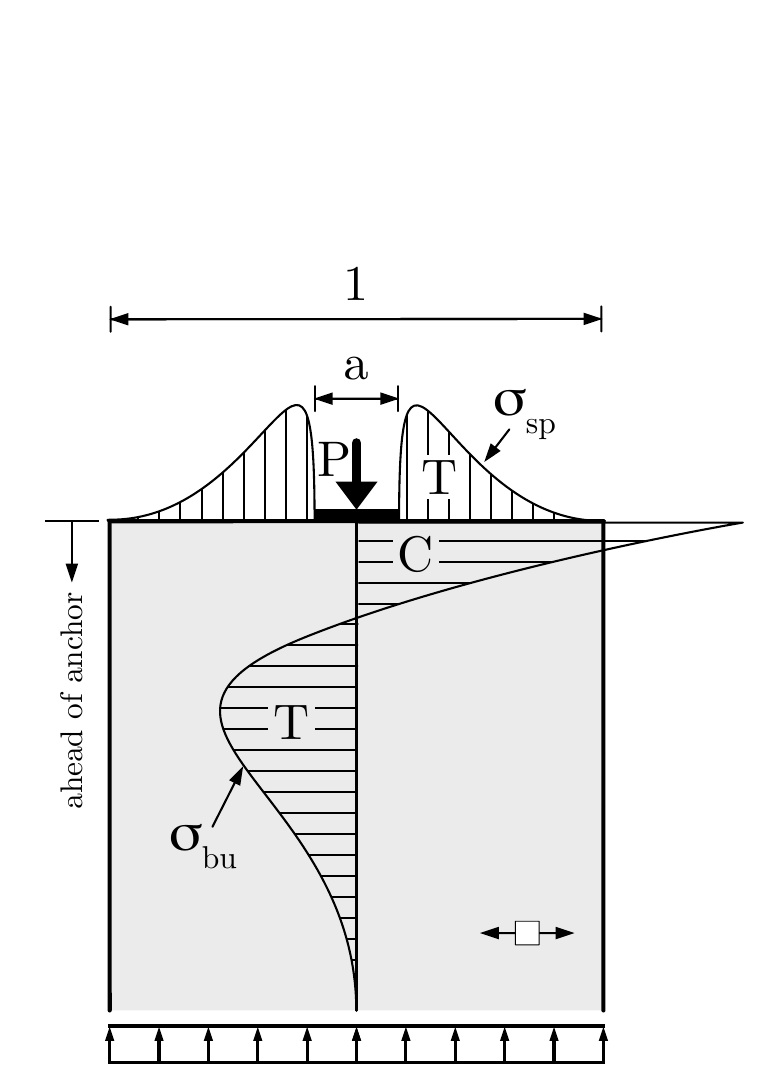}
    \caption{Transverse stresses at an end block.}
    \label{fig:stress_state_end}
  \end{subfigure}
  \caption{Elastic stress state in the anchorage zone for the two anchor
    positions. Adapted from
    \textcite{Wollmann1992} and \textcite{Guyon1974}.}
  \label{fig:stress_states}
\end{figure}

\subsection{Tie-back tension (d)}
\label{app:tieback}
Following \textcite{Wollmann1992}, the tie-back tension behind the plate
of an intermediate anchorage is
\begin{equation}
  \sigma_\mathrm{ia} = \frac{p_0}{2}\left(1-\frac{a}{c+c'}\right)< f_\mathrm{ct}\,, \qquad
  T_\mathrm{ia} = \frac{P}{2}\left(1-\frac{a}{c+c'}\right),
  \label{eq:sigma_ia}
\end{equation}
with $c + c'$ being the dimension in the direction of the anchorage considered,
Figure~\ref{fig:stress_state_intermediate}.
\textcite{Wollmann1992} derived
Eq.~\eqref{eq:sigma_ia} from plane-stress parameter studies of a slab
with a fully embedded anchor. 
The tied-back portion grows with the room for dispersion, so the direction with the larger block dimension governs, and Eq.~\eqref{eq:sigma_ia} evaluated on the full section bounds every anchor position from above \parencite{Wollmann1992} and for the compact plate the resultant is
referred to the plate area, $p_0 = P/A_0$, which bounds the tension
likewise. The recessed pocket of JOINT 1 and 3 interrupts the direct tension path through the anchor plane and the tied-back force is carried around the pocket through
the remaining concrete volume. The embedded case of \textcite{Wollmann1992} additionally develops a local transverse
tension $\sigma_\mathrm{bt} = 0.17\,p$ immediately behind the plate, see
Figure~\ref{fig:stress_state_intermediate}. However, this tensile stress does not develop at the recessed anchorages of JOINT~1 and JOINT~2, since no concrete lies behind the anchorage plate.

\subsection{Edge tension (e)}
\label{app:edge}
Following \textcite{Breen1994}, the edge tension for anchor positions outside the kern is
\begin{equation}
  \sigma_\mathrm{ed} = \left(6e_p - 1\right) p < f_\mathrm{ct}\,,
  \label{eq:sigma_ed}
\end{equation}
with $e_p$ being the eccentricity of the anchor from the centroid of the block.

\section{Additional details on interface failure}
\label{app:interface}

\subsection{Influence of external moment}
\label{app:mext}
\begin{figure}[b!]
  \centering
  \begin{subfigure}[t]{0.325\linewidth}
    \centering
    \includegraphics[width=\linewidth]{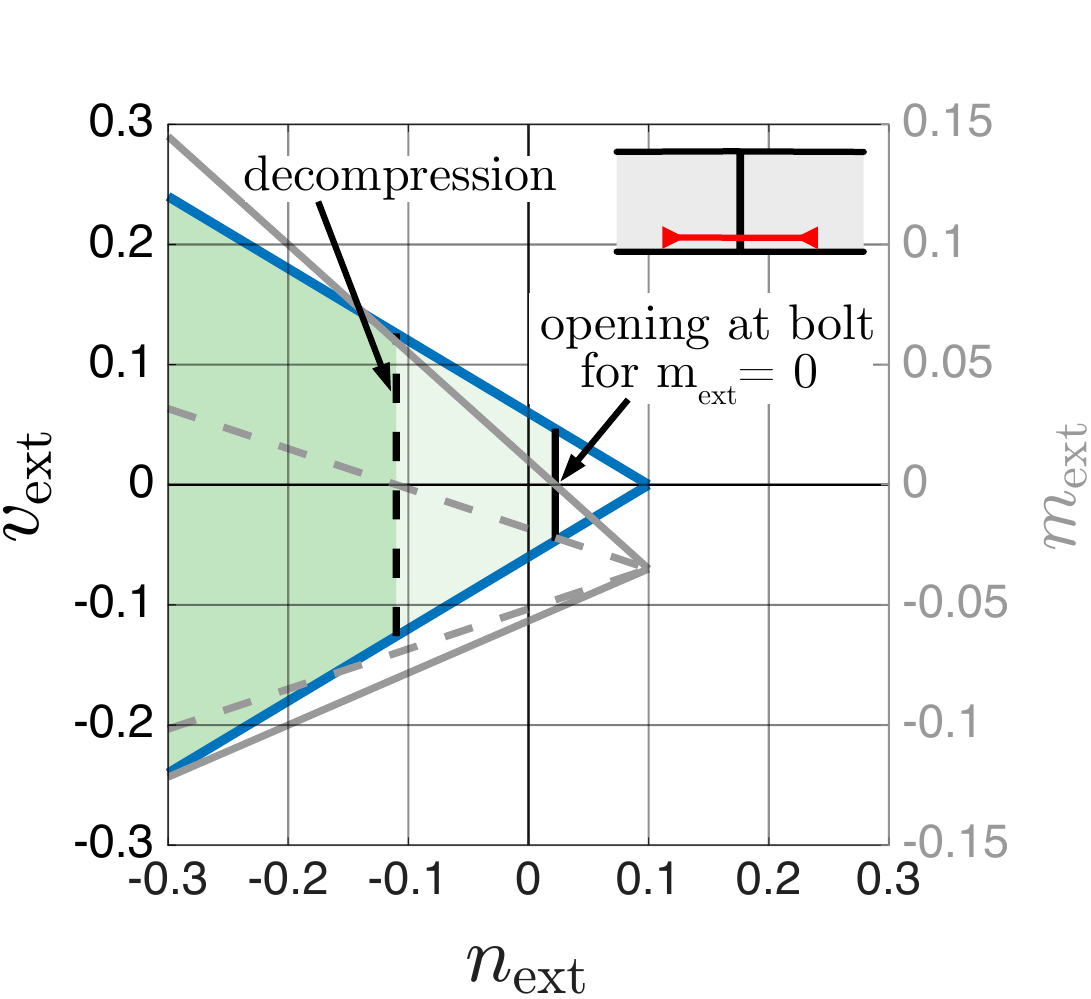}
    \caption{JOINT 1, $e_{\mathrm{p}} = 0.35$.}
    \label{fig:contact_limit_C1}
  \end{subfigure}\hfill
  \begin{subfigure}[t]{0.325\linewidth}
    \centering
    \includegraphics[width=\linewidth]{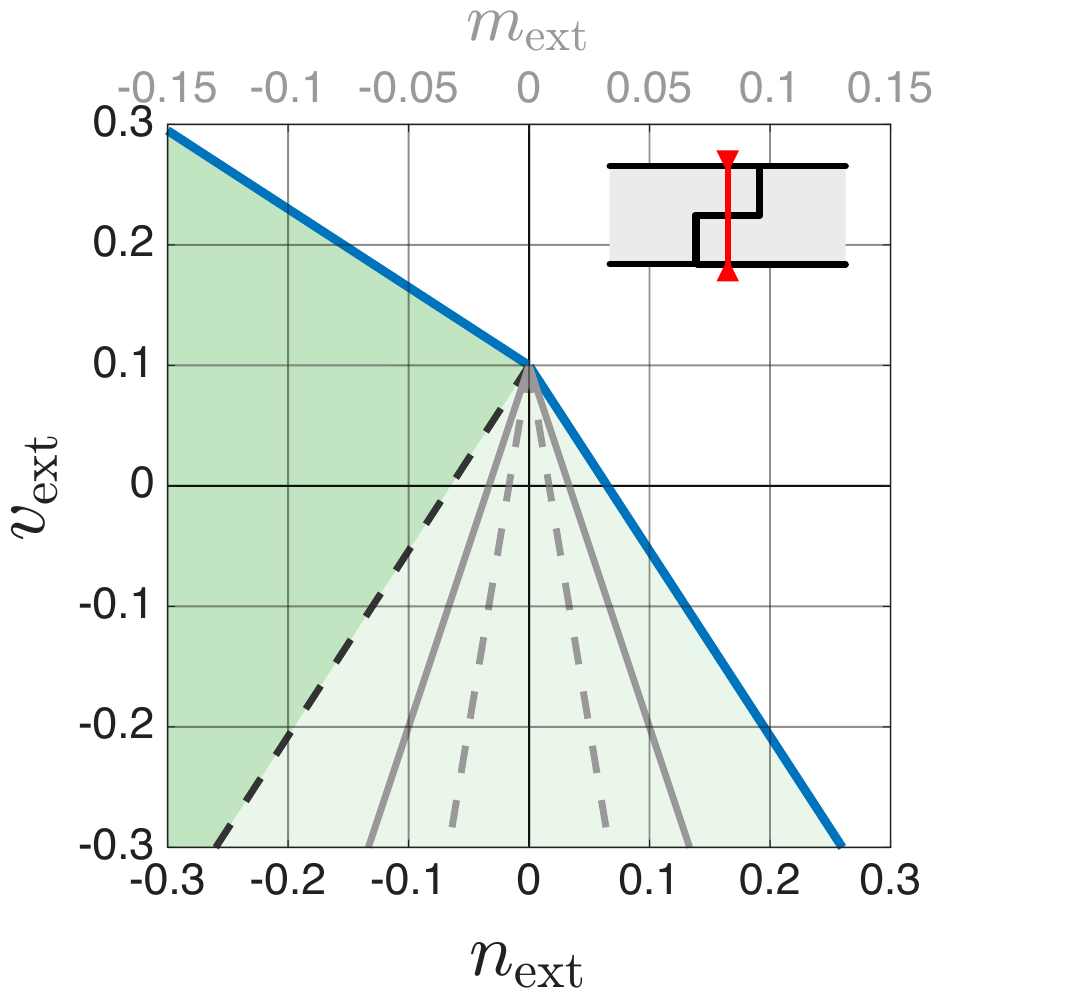}
    \caption{JOINT 2}
    \label{fig:contact_limit_C2}
  \end{subfigure}
  \begin{subfigure}[t]{0.325\linewidth}
    \centering
    \includegraphics[width=\linewidth]{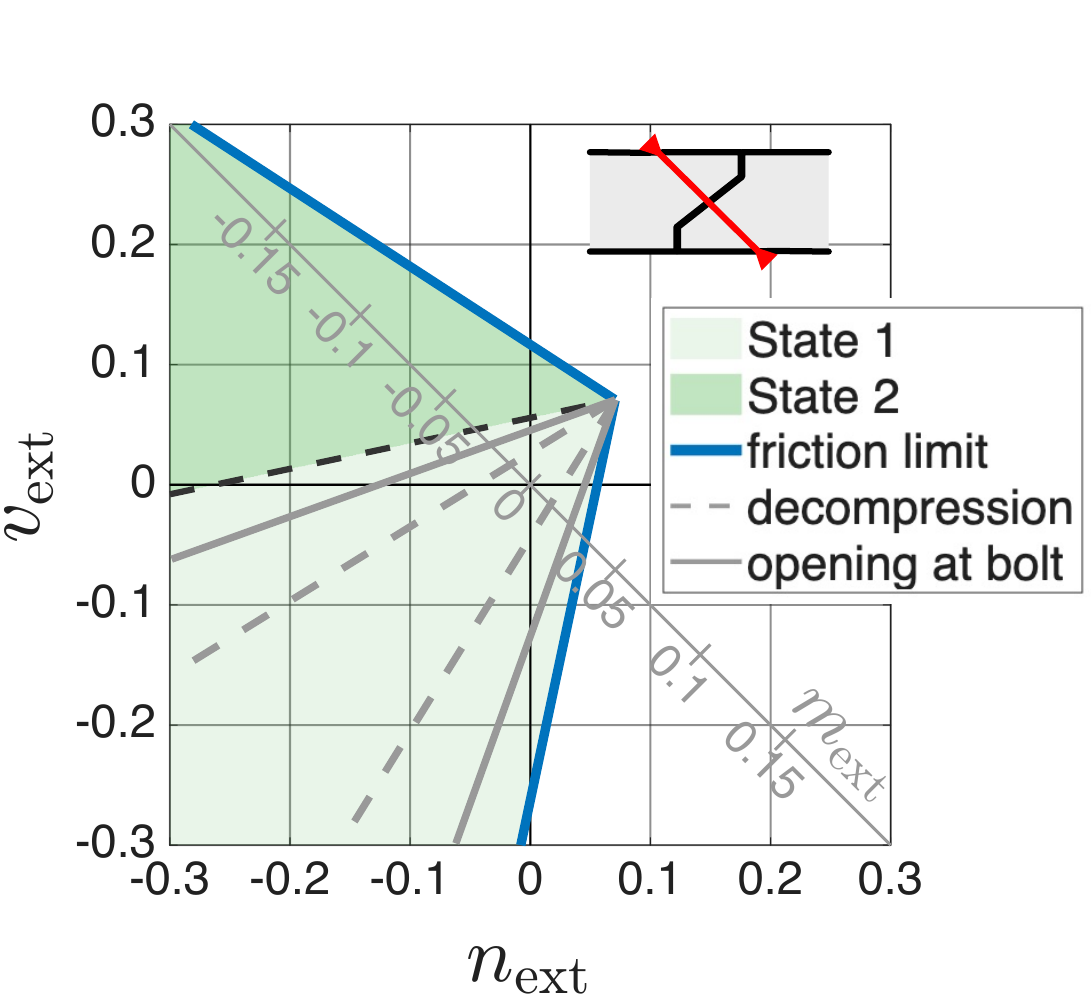}
    \caption{JOINT 3.}
    \label{fig:contact_limit_C3}
  \end{subfigure}
  \caption{Contact limit for interactions of $n_\mathrm{ext}$,
  $v_\mathrm{ext}$ and $m_\mathrm{ext}$, exemplary shown for $p=0.1$ .}
  \label{fig:app_mext}
\end{figure}

Figure~\ref{fig:app_mext} shows the contact limit for interactions of $n_\mathrm{ext}$, $v_\mathrm{ext}$ and $m_\mathrm{ext}$. At JOINT~1 the external moment adds to the moment of the eccentric prestress, Eq.~\eqref{eq:mI}. At JOINT~2 and JOINT~3 the post-tensioning member is positioned concentric on face I, so $m_{\mathrm{I}} = m_{\mathrm{ext}}$.

\subsection{Contact forces in state 2}
\label{app:contact}
In state 2 at JOINT 2 and 3 the contact forces of face~II follow from one of two
configurations,
\begin{equation}\label{eq:state2_alloc}
  \begin{aligned}
    &\text{face I in contact:}
    && n_{\mathrm{II}} =
      -\frac{t_{\mathrm{I}} + \mu\,n_{\mathrm{I}}}
            {\cos\alpha - \mu\sin\alpha} ,
    && t_{\mathrm{II}} = 0 , \\[6pt]
    &\text{face I open:}
    && n_{\mathrm{II}} = n_{\mathrm{ext}} - p\cos\beta ,
    && t_{\mathrm{II}} = v_{\mathrm{ext}} - p\sin\beta .
  \end{aligned}
\end{equation}
While face~I is in contact, face~II transfers by direct bearing the force
$n_{\mathrm{II}}$ by which the resultant force exceeds the friction limit of face~I. Face~I remains at this limit with the normal component
$n_{\mathrm{I}} - n_{\mathrm{II}}\sin\alpha$ and the traction component
$t_{\mathrm{I}} + n_{\mathrm{II}}\cos\alpha$. Face~I opens when its normal component reaches zero. Face~II then transfers all contact forces. Only in this configuration, face~II transfers a traction component.

\subsection{Verification} 
\label{app:verify}
At every location the normal stress $\sigma$ and the traction stress $\tau$ are transformed to the principal stresses
\begin{equation}\label{eq:principal}
  \sigma_{1,2} = \frac{\sigma}{2}
    \pm \sqrt{\Bigl(\frac{\sigma}{2}\Bigr)^{2} + \tau^{2}} ,
\end{equation}
which are compared with the biaxial failure envelope of Kupfer \cite{Kupfer1973}.
\printbibliography

\end{document}